\documentclass[aps,prd,twocolumn,superscriptaddress,nofootinbib,preprintnumbers]{revtex4-1}
\def\3x2pt{3$\times$2pt}
\def\5x2pt{5$\times$2pt}
\def\6x2pt{6$\times$2pt}

\newcommand{\kcmb}{\kappa_{\rm CMB}}

\newcommand{\metacal}{\textsc{Metacalibration}}

\newcommand{\redmagic}{\textsc{redMaGiC}}
\newcommand{\maglim}{\textsc{MagLim}}

\usepackage[utf8]{inputenc}
\usepackage[T1]{fontenc}
\usepackage{graphicx}% Include figure files
\usepackage{dcolumn}% Align table columns on decimal point
\usepackage{bm}% bold math
\usepackage{hyperref}% add hypertext capabilities
\usepackage{amsmath}	% Advanced maths commands
\usepackage{amssymb}	% Extra maths symbols
\usepackage{slashed}
\usepackage[normalem]{ulem}
\usepackage[dvipsnames]{xcolor}
\usepackage{xspace}
\usepackage{multirow}
\usepackage{booktabs}
\usepackage{comment}
\usepackage{lineno}
\RequirePackage{ifxetex}
\def\aj{AJ}%
\def\apj{ApJ}%
\def\apjs{ApJS}%
\def\aap{A\&A}%
\def\mnras{MNRAS}%
\def\prd{Phys.~Rev.~D}%
\def\pasp{PASP}%
\def\pasj{PASJ}%
\def\jcap{JCAP}%
\defcitealias{y3-nkgkmethods}{\textsc{Paper I}}
\defcitealias{y3-5x2kp}{\textsc{Paper III}}

\newcommand{\be}{\begin{equation}}
\newcommand{\ee}{\end{equation}}
\newcommand{\ba}{\begin{eqnarray}}
\newcommand{\ea}{\end{eqnarray}}

\newcommand{\nside}{\ifmmode N_{\mathrm{side}}\else $N_{\mathrm{side}}$\fi}
\newcommand{\npix}{\ifmmode n_{\mathrm{pix}}\else $n_{\mathrm{pix}}$\fi}
\newcommand{\Npix}{\ifmmode N_{\mathrm{pix}}\else $n_{\mathrm{pix}}$\fi}
\newcommand{\lmin}{\ifmmode \ell_{\mathrm{min}}\else $\ell_{\mathrm{min}}$\fi}
\newcommand{\lmax}{\ifmmode \ell_{\mathrm{max}}\else $\ell_{\mathrm{max}}$\fi}

\newcommand{\HEALPIX}{\textsc {Healpix}}

\newcommand{\mr}[1]{\mathrm{#1}}

\usepackage[utf8]{inputenc}
\usepackage{newunicodechar,graphicx}
\DeclareRobustCommand{\okina}{%
  \raisebox{\dimexpr\fontcharht\font`A-\height}{%
    \scalebox{0.8}{`}%
  }%
}
\newunicodechar{ʻ}{\okina}
\newcommand*\hawaii{Hawai\okina{}i}

\defcitealias{5x2methods}{B18}
\defcitealias{NKpaper}{O18a}
\defcitealias{GKpaper}{O18b}
\defcitealias{Krause:2017}{K17}
\defcitealias{Melchior:2017}{M17}
\defcitealias{Omori:2017}{O17}
\defcitealias{DESy1:2017}{DES-Y1-3x2}

\newcommand{\nk}{$\langle \delta_g \kappa_{\rm CMB}\rangle $}

\newcommand{\gk}{$\langle \gamma_{\rm t}\kappa_{\rm CMB}\rangle$}
\newcommand{\fivetwo}{$5\! \times\! 2 {\rm pt}$}
\newcommand{\threetwo}{$3\! \times\! 2 {\rm pt}$}

\newcommand{\nkgk}{$\langle \delta_g \kappa_{\rm CMB}\rangle + \langle \gamma_{\rm t}\kappa_{\rm CMB}\rangle$}

\begin{document}
\preprint{DES-2021-0648}
\preprint{FERMILAB-PUB-22-098-PPD}

\title[Cross-correlation of DES and CMB lensing]
{Joint analysis of DES Year 3 data and CMB lensing from SPT and Planck II:\\ Cross-correlation measurements and cosmological constraints}

% Author list file generated with: mkauthlist 1.2.4 
% mkauthlist -j prd --sort --aux order.csv DES-2021-0648_author_list.csv DES-2021-0648_author_list_part1.tex 

\author{C.~Chang}
\affiliation{Department of Astronomy and Astrophysics, University of Chicago, Chicago, IL 60637, USA}
\affiliation{Kavli Institute for Cosmological Physics, University of Chicago, Chicago, IL 60637, USA}
\author{Y.~Omori}
\affiliation{Department of Astronomy and Astrophysics, University of Chicago, Chicago, IL 60637, USA}
\affiliation{Kavli Institute for Cosmological Physics, University of Chicago, Chicago, IL 60637, USA}
\affiliation{Department of Physics, Stanford University, 382 Via Pueblo Mall, Stanford, CA 94305, USA}
\affiliation{Kavli Institute for Particle Astrophysics \& Cosmology, P. O. Box 2450, Stanford University, Stanford, CA 94305, USA}
\author{E.~J.~Baxter}
\affiliation{Institute for Astronomy, University of \hawaii, 2680 Woodlawn Drive, Honolulu, HI 96822, USA}
\author{C.~Doux}
\affiliation{Department of Physics and Astronomy, University of Pennsylvania, Philadelphia, PA 19104, USA}
\author{A.~Choi}
\affiliation{California Institute of Technology, 1200 East California Blvd, MC 249-17, Pasadena, CA 91125, USA}
\author{S.~Pandey}
\affiliation{Department of Physics and Astronomy, University of Pennsylvania, Philadelphia, PA 19104, USA}

\author{A.~Alarcon}
\affiliation{Argonne National Laboratory, 9700 South Cass Avenue, Lemont, IL 60439, USA}
\author{O.~Alves}
\affiliation{Department of Physics, University of Michigan, Ann Arbor, MI 48109, USA}
\affiliation{Laborat\'orio Interinstitucional de e-Astronomia - LIneA, Rua Gal. Jos\'e Cristino 77, Rio de Janeiro, RJ - 20921-400, Brazil}
\author{A.~Amon}
\affiliation{Kavli Institute for Particle Astrophysics \& Cosmology, P. O. Box 2450, Stanford University, Stanford, CA 94305, USA}
\author{F.~Andrade-Oliveira}
\affiliation{Department of Physics, University of Michigan, Ann Arbor, MI 48109, USA}
% ...

\collaboration{DES \& SPT Collaborations}

\date{Last updated \today}

\label{firstpage}

\begin{abstract}
Cross-correlations of galaxy positions and galaxy shears with maps of gravitational lensing of the cosmic microwave background (CMB) are sensitive to the distribution of large-scale structure in the Universe. Such cross-correlations are also expected to be immune to some of the systematic effects that complicate correlation measurements internal to galaxy surveys. We present measurements and modeling of the cross-correlations between galaxy positions and galaxy lensing measured in the first three years of data from the Dark Energy Survey with CMB lensing maps derived from a combination of data from the 2500 deg$^2$ SPT-SZ survey conducted with the South Pole Telescope and full-sky data from the {\it Planck} satellite. The CMB lensing maps used in this analysis have been constructed in a way that minimizes biases from the thermal Sunyaev Zel'dovich effect, making them well suited for cross-correlation studies. The total signal-to-noise of the cross-correlation measurements is 23.9 (25.7) when using a choice of angular scales optimized for a linear (nonlinear) galaxy bias model. We use the cross-correlation measurements to obtain constraints on cosmological parameters. For our fiducial galaxy sample, which consist of four bins of magnitude-selected galaxies, we find constraints of $\Omega_{\rm m} = 0.272^{+0.032}_{-0.052}$ and $S_{8} \equiv \sigma_8 \sqrt{\Omega_{\rm m}/0.3}= 0.736^{+0.032}_{-0.028}$ ($\Omega_{\rm m} = 0.245^{+0.026}_{-0.044}$ and $S_{8} = 0.734^{+0.035}_{-0.028}$) when assuming linear (nonlinear) galaxy bias in our modeling. Considering only the cross-correlation of galaxy shear with CMB lensing, we find $\Omega_{\rm m} = 0.270^{+0.043}_{-0.061}$ and $S_{8} = 0.740^{+0.034}_{-0.029}$. Our constraints on $S_8$ are consistent with recent cosmic shear measurements, but lower than the values preferred by primary CMB measurements from {\it Planck}. 

\end{abstract}

\maketitle

\section{Introduction}
\label{sec:intro}

Significant progress has been made recently in using cross-correlations between galaxy imaging and cosmic microwave background (CMB) surveys to constrain cosmological parameters. These developments have come naturally as ongoing galaxy and CMB surveys collect increasingly sensitive data across larger and larger overlapping areas of the sky. The Dark Energy Survey \citep[DES,][]{Flaugher2015} is the largest galaxy weak lensing survey today, covering $\sim 5000$ deg$^{2}$ of sky that is mostly in the southern hemisphere. By design, the DES footprint overlaps with high-resolution CMB observations from the South Pole Telescope \citep[SPT,][]{SPT}, enabling a large number of cross-correlation analyses \cite{soergel2016,Kirk2016,Giannantonio2016,Baxter:2016,Baxter2018,Prat2019,Omori19a,Omori19b,y15x2,Costanzi2021}. 

Although CMB photons originate from the high-redshift Universe, their trajectories are deflected by low-redshift structures as a result of gravitational lensing -- these are the same structures traced by the distributions of galaxies and the galaxy weak lensing signal measured in optical galaxy surveys.  Cross-correlating CMB lensing with galaxy surveys therefore allows us to extract 
information stored in the large-scale structure. 

In this work, we analyze both \nk{}, the cross correlation of the galaxy density field $\delta_g$ and the CMB weak lensing convergence field $\kappa_{\rm CMB}$, and \gk{}\footnote{The `t' subscript denotes the tangential component of shear, which will be discussed in Section~\ref{sec:estimator}}, the cross correlation of the galaxy weak lensing shear field $\gamma$ and $\kappa_{\rm CMB}$. Notably, these two two-point functions correlate measurements from very different types of surveys (galaxy surveys in the optical and CMB surveys in the millimeter), and are therefore expected to be very robust to systematic biases impacting only one type of survey.  Furthermore, CMB lensing is sensitive to a broad range of redshift, with peak sensitivity at redshift $z \sim 2$; galaxy lensing, on the other hand, is sensitive to structure at $z \lesssim 1$ for current surveys. As a result, the CMB lensing cross-correlation functions, \nkgk{}, are expected to increase in signal-to-noise relative to galaxy lensing correlations as one considers galaxy samples that extend to higher redshift.

Our analysis relies on the first three years (Y3) of galaxy observations from DES and a CMB lensing map constructed using data from the 2500 deg$^2$ SPT-SZ survey \citep{Omori:2017} and {\it Planck} \citep{Planck2020}.
The combined signal-to-noise of the \nkgk{} measurements used in the present cosmological analysis is roughly a factor two larger than in the earlier DES+SPT results presented in \cite{y15x2}, which used first year (Y1) DES data. This large improvement in signal-to-noise derives from two main advancements:
\begin{enumerate}
\item We have adopted a different methodology in constructing the CMB lensing map, which results in much lower contamination from the thermal Sunyaev Zel'dovich (tSZ) effect, allowing small-scale information to be used in the cosmological analysis. This methodology is described in \cite{y3-nkgkmethods}.
\item Data from DES Y3 covers an area approximately three times larger than DES Y1 and is slightly deeper. 
\end{enumerate}
Along with the significant increase in signal-to-noise, we have also updated our models for the correlation functions to include a number of improvements following \cite{y3-3x2ptkp}. These include an improved treatment of galaxy intrinsic alignments, inclusion of magnification effects on the lens galaxy density, and application of the so-called lensing ratio likelihood described in \cite{y3-shearratio}.

The analysis presented here is the second of a series of three papers: In \cite{y3-nkgkmethods} (\textsc{Paper I}) we describe the construction of the combined, tSZ-cleaned SPT+{\it Planck} CMB lensing map and the methodology for the cosmological analysis. In this paper (\textsc{Paper II}), we present the data measurements of the cross-correlation probes \nkgk{}, a series of diagnostic tests, and cosmological constraints from this cross-correlation combination. In \cite{y3-5x2kp} (\textsc{Paper III}), we will present the joint cosmological constraints from \nkgk{} and the DES-only \threetwo{} probes\footnote{The \threetwo{} probes refer to a combination of three two-point functions of the galaxy density field $\delta_{g}$ and the weak lensing shear field $\gamma$: galaxy clustering $\langle \delta_{g} \delta_{g} \rangle$, galaxy-galaxy lensing $\langle \delta_{g} \gamma_{\rm t} \rangle$ and cosmic shear $\langle \gamma \gamma \rangle$.}, and tests of consistency between the two, as well as constraints from a joint analysis with the CMB lensing auto-spectrum. 

Similar analyses have recently been carried out using different galaxy imaging surveys and CMB data. \cite{Marques2020} studied the  cross-correlation of the galaxy weak lensing from the Hyper Suprime-Cam Subaru Strategic Program Survey \citep[HSC-SSP,][]{HSC} and the {\it Planck} lensing map \citep{Planck:cmblensing}; \cite{Namikawa2019} used the same HSC galaxy weak lensing measurement to cross-correlate with CMB lensing from the POLARBEAR experiment \cite{Kermish2012}; \cite{Robertson2021} cross-correlated galaxy weak lensing from the Kilo-Degree Survey \citep[KiDS,][]{KIDS} and the CMB lensing map from the Atacama Cosmology Telescope \citep[ACT,][]{ACT}; and \cite{Krolewski2021} cross-correlated the galaxy density measured in unWISE data \cite{Schlafly2019} with {\it Planck} CMB lensing. Compared to these previous studies, in addition to the new datasets, this paper is unique in that we combine \nk{} and \gk{}.  Moreover, our analysis uses the same modeling choices and analysis framework as in \cite{y3-3x2ptkp}, making it easy to compare and combine our results later (i.e. \citetalias{y3-5x2kp}).

The structure of the paper is as follows. In Section~\ref{sec:model} we briefly review the formalism of our model for the two cross-correlation functions and the parameter inference pipeline (more details can be found in \citetalias{y3-nkgkmethods}). In Section~\ref{sec:data} we review the data products used in this analysis. In Section~\ref{sec:estimator} we introduce the estimators we use for the correlation functions. In Section~\ref{sec:blinding} we describe out blinding procedure and unblinding criteria. In Section~\ref{sec:results} we present constraints on cosmological parameters as well as relevant nuisance parameters when fitting to the cross-correlation functions. Finally we conclude in Section~\ref{sec:discussion}.

\section{Modelling and inference}
\label{sec:model}

We follow the theoretical formalism laid out in \citetalias{y3-nkgkmethods} and \cite{y3-generalmethods} for this work. Here, we summarize only the main equations relevant to this paper. Following standard convention, we refer to the galaxies used to measure $\delta_g$ as \textit{lens} galaxies, and the galaxies used to measure $\gamma$ as \textit{source} galaxies.\\ 

\noindent {\bf Angular power spectra:} Using the Limber approximation\footnote{In \cite{Fang2020}, the authors showed that at DES Y3 accuracy, the Limber approximation is sufficient for galaxy-galaxy lensing and cosmic shear but insufficient for galaxy clustering. Given the primary probe in this work, \nkgk{}, are at much lower signal-to-noise than galaxy-galaxy lensing and cosmic shear, we expect that Limber approximation is still a valid choice.} \cite{Limber:1953}, the cross-spectra between CMB lensing convergence and galaxy density/shear can be related to the matter power spectrum via:
\begin{align}
C&^{\kappa_{\rm CMB} X^{i}}(\ell) = \notag \\
&\int d\chi \frac{q_{\kappa_{\rm CMB}}(\chi)q^i_{X} (\chi)}{\chi^2} P_{\rm NL} \left( \frac{\ell+1/2}{\chi}, z(\chi)\right),
\label{eq:Cl_basic}
\end{align}
where $X \in \{\delta_g, \gamma \}$, $i$ labels the redshift bin, $P_{\rm NL}(k,z)$ is the non-linear matter power spectrum, which we compute using \texttt{CAMB} and \texttt{Halofit} \citep{camb,Takahashi:2012}, and $\chi$ is the comoving distance to redshift $z$. The weighting functions, $q_{X}(\chi)$, describe how the different probes respond to large-scale structure at different distances, and are given by 
\begin{equation}
q_{\kappa_{\rm CMB}} (\chi) = \frac{3 \Omega_{\rm m} H_0^2}{2c^2}\frac{\chi}{a(\chi)}  \frac{\chi^* - \chi}{\chi^*}, \label{eq:weight_cmbkappa}
\end{equation}
\begin{equation}
q^i_{\delta_g}(\chi) = b^i(k,z(\chi)) n_{\delta_{g}}^i(z(\chi)) \frac{dz}{d\chi}
\label{eq:q_deltag}
\end{equation}
\begin{equation}
q^{i}_{\gamma}(\chi) = \frac{3H_0^2 \Omega_{\rm m}}{2c^2} \frac{\chi}{a(\chi)} \int_{\chi}^{\chi_h} d\chi' 
n_{\gamma}^i(z(\chi')) 
\frac{dz}{d\chi'}
\frac{\chi' -\chi}{\chi'}, 
\label{eq:q_gamma}
\end{equation}
where $H_0$ and $\Omega_{\rm m}$ are the Hubble constant and matter density parameters, respectively, $a(\chi)$ is the scale factor corresponding to comoving distance $\chi$, $\chi^*$ denotes the comoving distance to the surface of last scattering, $b(k,z)$ is galaxy bias as a function of scale ($k$) and redshift, and  $n^i_{{\delta_{g}}/\gamma}(z)$ are the normalized redshift distributions of the lens/source galaxies in bin $i$.\\ 

\noindent \textbf{Correlation functions:} The angular-space correlation functions are then computed via
\begin{align}
w^{\delta_g^i \kappa_{\rm CMB}}&(\theta) =\notag \\
&\sum_{\ell} \frac{2\ell + 1}{4\pi} F(\ell) P_{\ell} (\cos(\theta)) C^{\delta_g^i \kappa_{\rm CMB}}(\ell), 
\label{eq:wnk}
\end{align}
\begin{align}
w^{\gamma_{t}^{i} \kappa_{\rm CMB}} &(\theta)= \notag \\ 
&\sum_{\ell}\frac{2\ell+1}{4\pi\ell(\ell+1)}F(\ell) P_{\ell}^{2}(\cos\theta)C^{\kappa^{i}_{\gamma}\kappa_{\rm CMB}}(\ell),
\label{eq:wgk}
\end{align}
where $P_{\ell}$ and $P^2_{\ell}$ are the $\ell$th order Legendre polynomial and associated Legendre polynomial, respectively, and $F(\ell)$ describes filtering applied to the $\kappa_{\rm CMB}$ maps.   For correlations with the $\kappa_{\rm CMB}$ maps, we set $F(\ell)= B(\ell) H(\ell - \ell_{\rm min}) H(\ell_{\rm max} - \ell)$, where $H(\ell)$ is a step function and $B(\ell) = \exp (-0.5\ell(\ell + 1)\sigma^2)$ with $\sigma \equiv \theta_{\rm FWHM}/\sqrt{8 \ln 2}$, and $\theta_{\rm FWHM}$ describes the beam applied to the CMB lensing maps (see discussion of $\ell_{\rm min}$, $\ell_{\rm max}$, and $\theta_{\rm FWHM}$ choices in Section~\ref{sec:data}, and further discussion in \citetalias{y3-nkgkmethods}). \\ 

\noindent {\bf Galaxy bias:} We consider two models for the galaxy bias $b(k,z)$. Our fiducial choice is a linear bias model where $b(k,z)=b^{i}$ is not a function of scale and is assumed to be a free parameter for each tomographic bin $i$. The second bias model is an effective 1-loop model with renormalized nonlinear galaxy bias parameters: $b^{i}_1$ (linear bias), $b^{i}_2$ (local quadratic bias), $b^{i}_{s^2}$ (tidal quadratic bias) and $b^{i}_{\rm 3nl}$ (third-order non-local bias). The latter two parameters can be derived from $b^{i}_1$, making the total number of free parameters for this bias model two per tomographic bin $i$ \cite{y3-2x2ptbiasmodelling}. \\   

\noindent {\bf Intrinsic alignment (IA):} Galaxy shapes can be intrinsically aligned as a result of nearby galaxies evolving in a common tidal field. IA modifies the observed lensing signal. We adopt the five-parameter ($a_1$,$\eta_1$,$a_2$,$\eta_2$,$b_{\rm ta}$) tidal alignment tidal torquing model (TATT) of \cite{blazek2019} to describe galaxy IA. $a_1$ and $\eta_1$ characterize the amplitude and redshift dependence of the tidal alignment; $a_2$ and $\eta_2$ characterize the amplitude and redshift dependence of the tidal torquing effect; $b_{\rm ta}$ accounts for the fact that our measurement is weighted by the observed galaxy counts. In Section~\ref{sec:SR_IA}, we will also compare our results using a simpler IA model, the nonlinear alignment model \citep[NLA,][]{bridle07}. The TATT model is equivalent to the NLA model in the limit that $a_2=\eta_2=b_{\rm ta}=0$. \\

\noindent {\bf Impact of lensing magnification on lens galaxy density:} Foreground structure modulates the observed galaxy density as a result of gravitational magnification. The effect of magnification can be modeled by modifying Equation~\ref{eq:q_deltag} to include the change in selection and geometric dilution quantified by the lensing bias coefficients $C_g^i$:
\begin{equation}
q^i_{\delta_{g}}(\chi) \rightarrow q^i_{\delta_{g}}(\chi) (1+C^{i}_{g}\kappa^{i}_g),
\label{eq:magnification}
\end{equation}
where $\kappa_g^i$ is the tomographic convergence field, as described in \cite{y3-generalmethods} and the values of $C_g^i$ are estimated in \citep{y3-2x2ptmagnification} and fixed to the values listed in Table~\ref{table:prior}.\\

\noindent {\bf Uncertainty in redshift distributions:} We model uncertainty in the redshift distributions of the source galaxies with shift parameters, $\Delta z^i$, defined such that for each redshift bin $i$,
    \begin{equation}
    \label{eq:nz}
    n^{i}(z) \rightarrow n^{i}(z-\Delta^{i}_{z}).
    \end{equation}    
For the lens sample, we additionally introduce a stretch parameter ($\sigma_{z}$) when modeling the redshift distribution, as 
motivated by \cite{y3-lenswz}:
    \begin{equation}
    n^{i}(z) \rightarrow \sigma_{z}^{i}n^{i}(\sigma_{z}^{i}[z-\langle z \rangle] + \langle z \rangle -\Delta^{i}_{z}),
    \end{equation}
where $\langle z \rangle$ is the mean redshift.    \\
    
\noindent {\bf Uncertainty in shear calibration:} We model uncertainty in the shear calibration with multiplicative factors defined such that the observed $C^{\kappa_{\rm CMB}\gamma}$ is modified by
\begin{equation}\label{eq:shear_bias}
C^{\kappa_{\rm CMB}\gamma^{i}}(\ell) \rightarrow (1+m^{i})C^{\kappa_{\rm CMB}\gamma^{i}}(\ell),
\end{equation}
where $m^{i}$ is the shear calibration bias for source bin $i$. \\

\noindent {\bf Lensing ratio (or shear ratio, SR):} The DES Y3 \threetwo{} analysis used a ratio of small-scale galaxy lensing measurements to provide additional information, particularly on source galaxy redshift biases and on IA parameters. These ratios are not expected to directly inform the cosmological constraints; they can, however, improve constraints via degeneracy breaking with nuisance parameters. The lensing ratios can therefore be considered as another form of systematic calibration, in a similar vein to, e.g., spectroscopic data used to calibrate redshifts, and image simulations used to calibrate shear biases.  In \citep{y3-shearratio}, it was demonstrated that the lensing ratio measurements are approximately independent of the \threetwo{} measurements, making it trivial to combine constraints from \threetwo{} and lensing ratios at the likelihood level. Unless otherwise mentioned, all our analyses will include the information from these lensing ratios. We investigate their impact in Section~\ref{sec:SR_IA}.\\

\noindent {\bf Angular scale cuts:} The theoretical model described above is uncertain on small scales due to uncertainty in our understanding of baryonic feedback and the galaxy-halo connection (or, nonlinear galaxy bias). We take the approach of only fitting the correlation functions on angular scales we can reliably model.  In \citetalias{y3-nkgkmethods} we determined the corresponding angular scale cuts by requiring the cosmological constraints to not be significantly biased when prescriptions for unmodeled effects are introduced.  In Figures~\ref{fig:davtavec_maglim} and~\ref{fig:wnk} the scale cuts are marked by the grey bands.\\

\noindent {\bf Parameter inference:} We assume a Gaussian likelihood\footnote{See e.g. \citep{Lin2020} for tests of the validity of this assumption in the context of cosmic shear, which would also apply here.} for the data vector of measured correlation functions, $\vec{d}$, given a model, $\vec{m}$, generated using the set of parameters $\vec{p}$:
\begin{align}
\ln \mathcal{L}&(\vec{d}|\vec{m}(\vec{p}))=  \notag \\
&-\frac{1}{2} \sum^N_{ij} \left(d_i - m_i(\vec{p})\right)^T \mathbf{C}^{-1}_{ij} \left(d_j - m_j(\vec{p}) \right),  
\end{align}
where the sums run over all of the $N$ elements in the data and model vectors. The posterior on the model parameters is then given by:
\begin{equation}
P(\vec{m}(\vec{p})|\vec{d}) \propto \mathcal{L}(\vec{d} | \vec{m}(\vec{p})) P_{\rm prior} (\vec{p}),
\end{equation}
where $P_{\rm prior}(\vec{p})$ is a prior on the model parameters. Our choice of priors is summarized in Table~\ref{table:prior}.

The covariance matrix used here consists of an analytical lognormal covariance combined with empirical noise estimation from simulations. The covariance has been extensively validated in \citetalias{y3-nkgkmethods}. In Appendix~\ref{sec:jk_cov} Figure~\ref{fig:cov_jk_vs_analytical} we show that the diagonal elements of our final analytic covariance are in excellent agreement with a covariance estimated from data using jackknife resampling.    

Our modeling and inference framework is built within the \texttt{CosmoSIS} package \cite{cosmosis} and is designed to be consistent with those developed as part of \cite{y3-3x2ptkp}. We generate parameter samples using the nested sampler \textsc{PolyChord} \cite{polychord}. 

\begin{table*}
\centering 
\begin{tabular}{cc}
\hline
Parameter & Prior  \\ \hline
$\Omega_{\rm{m}} $ & $\mathcal{U}[0.1, 0.9]$  \\
$A_{\rm{s}}\times 10^{9}$ & $\mathcal{U}[0.5, 5.0]$ \\
$n_{\rm{s}}$ & $\mathcal{U}[0.87, 1.07]$ \\
$\Omega_{\rm{b}}$ & $\mathcal{U}[0.03, 0.07]$  \\
$h$ & $\mathcal{U}[0.55, 0.91]$\\
$\Omega_{\nu}h^2 \times 10^{4} $& $\mathcal{U}[6.0, 64.4]$  \\  
\hline
$a_1$ & $\mathcal{U}[-5.0, 5.0]$ \\
$a_2$ & $\mathcal{U}[-5.0, 5.0]$\\
$\eta_1$ & $\mathcal{U}[-5.0, 5.0]$ \\
$\eta_2$ & $\mathcal{U}[-5.0, 5.0]$ \\
$b_{\rm{ta}}$ & $\mathcal{U}[0.0, 2.0]$ \\ 
\hline
\textsc{MagLim} &  \\
$b^{1\cdots 6}$  & $\mathcal{U}[0.8, 3.0]$ \\
$b_{1}^{1\cdots 6}$  & $\mathcal{U}[0.67, 3.0]$\\
$b_{2}^{1\cdots 6}$ & $\mathcal{U}[-4.2, 4.2]$ \\
$C_{\rm l}^{1\cdots 6}$ & $\delta(1.21)$, $\delta(1.15)$, $\delta(1.88)$, $\delta(1.97)$, \textcolor{gray}{$\delta(1.78)$, $\delta(2.48)$}  \\
$\Delta_z^{1...6} \times 10^{2}$ & $\mathcal{N}[-0.9, 0.7]$, $\mathcal{N}[-3.5, 1.1]$, $\mathcal{N}[-0.5, 0.6]$,  $\mathcal{N}[-0.7, 0.6]$, \textcolor{gray}{$\mathcal{N}[0.2, 0.7]$ ,  $\mathcal{N}[0.2, 0.8]$}  \\
$\sigma_{z}^{1...6}$ & $\mathcal{N}[0.98, 0.062]$, $\mathcal{N}[1.31, 0.093]$, $\mathcal{N}[0.87, 0.054]$, $\mathcal{N}[0.92,  0.05]$, \textcolor{gray}{$\mathcal{N}[1.08, 0.067]$, $\mathcal{N}[0.845, 0.073]$}  \\
 \hline
\textcolor{gray}{\redmagic{}} &  \\
\textcolor{gray}{$b^{1\cdots 5}$}  & \textcolor{gray}{$\mathcal{U}[0.8, 3.0]$} \\
\textcolor{gray}{$b_{1}^{1\cdots 5}$}  & \textcolor{gray}{$\mathcal{U}[0.67, 2.52]$} \\
\textcolor{gray}{$b_{2}^{1\cdots 5}$} & \textcolor{gray}{$\mathcal{U}[-3.5, 3.5]$} \\
\textcolor{gray}{$C_{\rm l}^{1\cdots 5}$} & \textcolor{gray}{$\delta(1.31)$, $\delta(-0.52)$, $\delta(0.34)$, $\delta(2.25)$, $\delta(1.97)$} \\ 
\textcolor{gray}{$\Delta_z^{1...5} \times 10^{2}$} & \textcolor{gray}{$\mathcal{N}[0.6, 0.4]$, $\mathcal{N}[0.1, 0.3]$, $\mathcal{N}[0.4, 0.3]$,}  \textcolor{gray}{$\mathcal{N}[-0.2, 0.5]$, $\mathcal{N}[-0.7, 1.0]$}  \\
\textcolor{gray}{$\sigma_{z}^{1...4}$} & \textcolor{gray}{$\delta(1)$, $\delta(1)$, $\delta(1)$, $\delta(1)$, $\mathcal{N}[1.23, 0.054]$}  \\
\hline
\textsc{MetaCalibration} &  \\
$m^{1...4} \times 10^{3}$ &  $\mathcal{N}[-6.0, 9.1]$, $\mathcal{N}[-20.0, 7.8]$,  $\mathcal{N}[-24.0, 7.6]$, $\mathcal{N}[-37.0, 7.6]$ \\
$\Delta_z^{1...4} \times 10^{-2}$ & $\mathcal{N}[0.0, 1.8]$, $\mathcal{N}[0.0, 1.5]$,  $\mathcal{N}[0.0, 1.1]$, $\mathcal{N}[0.0, 1.7]$   \\
\hline
\end{tabular}
\caption{Prior values for cosmological and nuisance parameters included in our model. For the priors, $\mathcal{U}[a,b]$ indicates a uniform prior between $a$ and $b$, while $\mathcal{N}[a,b]$ indicates a Gaussian prior with mean $a$ and standard deviation $b$. $\delta(a)$ is a Dirac Delta function at value $a$, which effectively means that the parameter is fixed at $a$. Note that the fiducial lens sample is the first 4 bins of the \maglim{} sample. The two high-redshift \maglim{} bins and the \redmagic{} sample are shown in grey to indicate they are not part of the fiducial analysis.}
\label{table:prior}
\end{table*}

\section{Data}
\label{sec:data}

\subsection{CMB lensing maps}
There are two major advances in the galaxy and CMB data used here relative to the DES Y1 and SPT analysis presented in \cite{Omori19a, Omori19b}. First, for the CMB map in the SPT footprint, we used the method developed in \cite{madhavacheril18} and described in \citetalias{y3-nkgkmethods} to remove contamination from the tSZ effect by combining data from SPT and {\it Planck}. Such contamination was one of the limiting factors in our Y1 analysis. Second, the DES Y3 data cover a significantly larger area on the sky than the DES Y1 data. Consequently, the DES Y3 footprint extends beyond the SPT footprint, necessitating the use of the {\it Planck}-only lensing map \citep{Planck2020} over part of the DES Y3 patch. As discussed in \citetalias{y3-nkgkmethods}, the different noise properties and filtering of the two lensing maps necessitates separate treatment throughout. The ``SPT+{\it Planck}'' lensing map, which overlaps with the DES footprint at $<-40$ degrees in declination, is filtered by $\ell_{\rm min}=8$, $\ell_{\rm max}=5000$ and a Gaussian smoothing of $\theta_{\rm FWHM}=6$ arcmin. This map is produced using the combination of 150 GHz data from the 2500 deg$^2$ SPT-SZ survey \citep[e.g.,][]{Omori:2017}, Planck 143 GHz data, and the tSZ-cleaned CMB Planck temperature map generated using the Spectral Matching Independent Component Analysis (SMICA) algorithm (i.e. the SMICA-noSZ map).  The ``{\it Planck}'' lensing map, which overlaps with the DES footprint at $>-39.5$ degrees in declination, is filtered by $\ell_{\rm min}=8$, $\ell_{\rm max}=3800$ and a Gaussian smoothing of $\theta_{\rm FWHM}=8$ arcmin is applied. This map is reconstructed using the {\it Planck} SMICA-noSZ temperature map alone. We leave a small 0.5 deg gap between the two lensing maps to reduced the correlation between structures on the boundaries. The resulting effective overlapping areas with DES are $1764\ {\rm deg}^{2}$ and $2156\ {\rm deg}^{2}$ respectively for the SPT+{\it Planck} and {\it Planck} patches respectively.

\subsection{The DES Y3 data products}

DES \citep{Flaugher2005} is a photometric survey in five broadband filters ($grizY$), with a footprint of nearly $5000 \; {\rm deg}^2$ of sky that is mostly in the southern hemisphere, imaging hundreds of millions of galaxies. It employs the 570-megapixel Dark Energy Camera \citep[DECam,][]{Flaugher2015} on the Cerro Tololo Inter-American Observatory (CTIO) 4m Blanco telescope in Chile. We use data from the first three years (Y3) of DES observations. The foundation of the various DES Y3 data products is the Y3 Gold catalog described in \cite{y3-gold}, which achieves S/N$\sim$10 for extended objects up to i$\sim$23.0 over an unmasked area of $4143 \; {\rm deg}^2$. In this work we use three galaxy samples: two lens samples for the galaxy density-CMB lensing correlation, \nk{}, and one source sample for the galaxy shear-CMB lensing correlation, \gk{}. We briefly describe each sample below. These samples are the same as those used in \cite{y3-3x2ptkp} and we direct the readers to a more detailed description of the samples therein. 

\begin{figure}
\begin{center}
\includegraphics[width=0.9\linewidth]{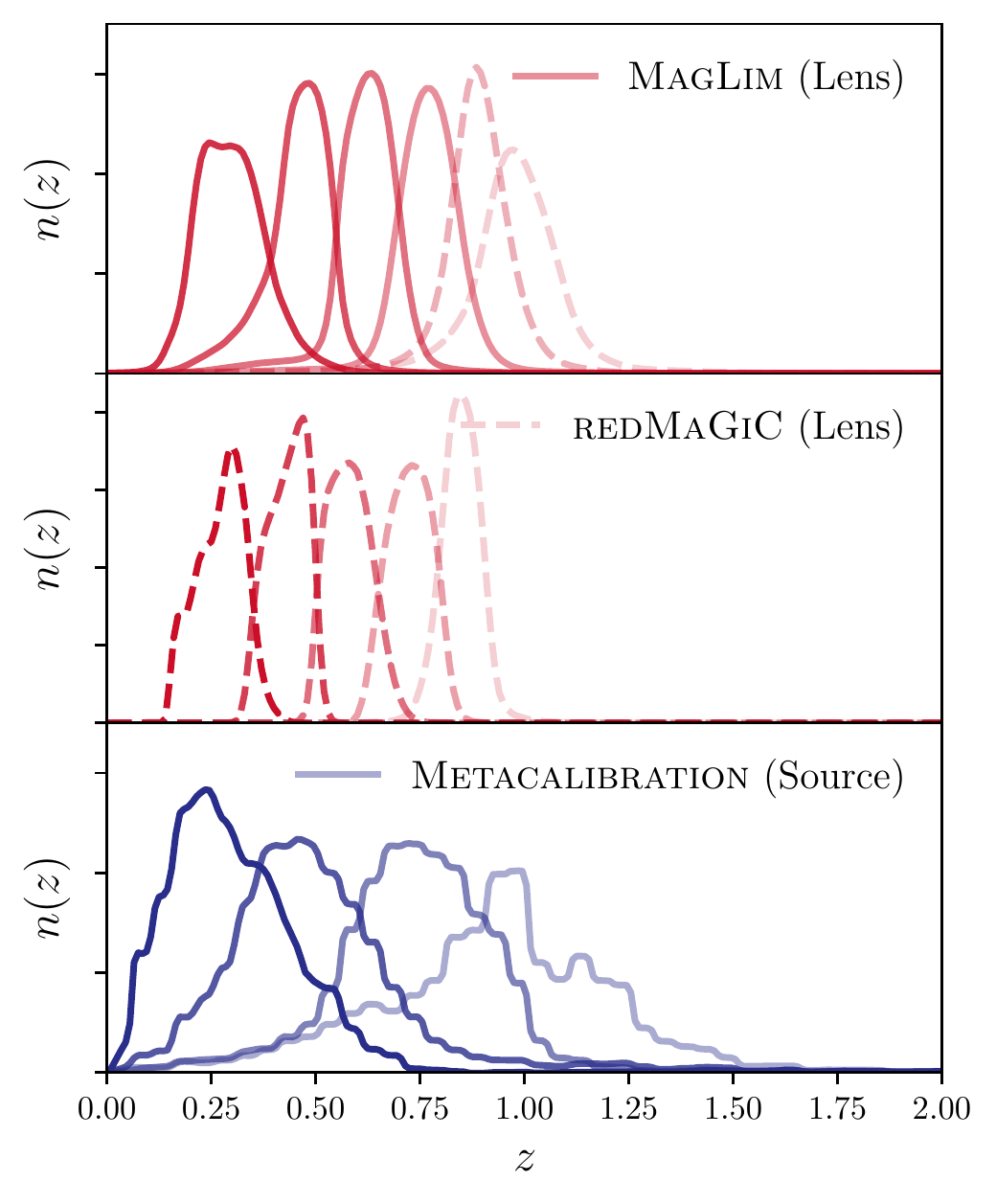}
\caption{Redshift distribution for the tomographic bins for the galaxy samples used in this work: the \maglim{} lens sample (top), the \redmagic{} lens sample (middle) and the \textsc{Metacal} source sample (bottom). The fiducial lens sample only uses the first four bins of the \maglim{} sample, or the solid lines. We perform tests with the non-fiducial samples (dashed lines) for diagnostic purposes.}
\label{fig:zdists}
\end{center}
\end{figure}

\subsubsection{Lens samples: \maglim{} and \redmagic{}}
\label{sec:lenses}

We will show results from two lens galaxy samples named \maglim{} and \redmagic{}. Following \cite{y3-3x2ptkp}, the first four bins of the \maglim{} sample  will constitute our fiducial sample, though we show results from the other bins and samples to help understand potential systematic effects in the DES galaxy selection. 

The \maglim{} sample consists of 10.7 million galaxies selected with a magnitude cut that evolves linearly with the photometric redshift estimate: $i < 4 z_{\rm phot} + 18$. z$_{\rm phot}$ is determined using the Directional Neighborhood Fitting algorithm \citep[DNF,][]{DeVicente2016}. \cite{y3-2x2maglimforecast} optimized the magnitude cut to balance the statistical power of the sample size and the accuracy of the photometric redshifts for cosmological constraints from galaxy clustering and galaxy-galaxy lensing. \maglim{} is divided into six tomographic bins.  The top panel of Figure~\ref{fig:zdists} shows the per-bin redshift distributions, which have been validated using cross-correlations with spectroscopic galaxies in \cite{y3-lenswz}.  Weights are derived to account for survey systematics, as described in \cite{y3-galaxyclustering}.

The \redmagic{} sample consists of 2.6 million luminous red galaxies (LRGs) with 
small photometric redshift errors \citep{Rozo2016redmagic}. \redmagic{} is constructed using a red sequence template calibrated via the \textsc{redMaPPer} algorithm \citep{rykoff2014,rykoff2016}. The lens galaxies are divided into five tomographic bins. 
The redshift distributions are shown in the middle panel of Figure~\ref{fig:zdists}. These distributions are estimated using draws from the redshift probability distribution functions of the individual \redmagic{} galaxies. As with \maglim{}, \cite{y3-lenswz} validates the redshift distributions, and \cite{y3-galaxyclustering} derives systematics weights. 

We note that in \cite{y3-3x2ptkp} the two high-redshift bins were excluded in \maglim{} due to poor fits in the \threetwo{} analysis, while the \redmagic{} sample was excluded due to an internal tension between galaxy-galaxy lensing and galaxy clustering. With the addition of CMB lensing cross-correlations, one of the aims of this work will be to shed light on potential systematic effects in the lens samples.   We briefly discuss this issue in Section~\ref{sec:Xlens} but there will be a more in-depth discussion in \citetalias{y3-5x2kp} when we combine with the \threetwo{} probes.

\subsubsection{Source sample: \metacal}
For the source sample, we use the DES Y3 shear catalog presented in \cite{y3-shapecatalog}, which contains over 100 million galaxies. The galaxy shapes are estimated using the $\metacal$ algorithm \citep{huff2017,sheldon2017}. The shear catalog has been thoroughly tested in \cite{y3-shapecatalog, y3-imagesims}. In \cite{y3-imagesims}, the authors used realistic image 
simulations to constrain the multiplicative bias of the shear estimate to be at most 2-3\%, primarily attributed to a shear-dependent detection bias coupled with object blending effects. The residual shear calibration biases are folded into the modeling pipeline and are listed in Table~\ref{table:prior}.

The source galaxies are divided into four tomographic bins based on the SOMPZ algorithm described in \citep{y3-sompz}, utilizing deep field data described in \cite{y3-deepfields} and image simulations described in \cite{y3-balrog}. The bottom panel of Figure~\ref{fig:zdists} shows the redshift distributions, which have been validated in \cite{y3-sourcewz} and \cite{y3-shearratio}.

\section{Correlation function estimators}
\label{sec:estimator}

Our estimator for the galaxy-CMB lensing correlation (Equation~\ref{eq:wnk}) is
\begin{align}
\langle \delta_{g} \kcmb(\theta_{\alpha}) \rangle = \langle \delta_{g} \kcmb(\theta_{\alpha}) \rangle_{0} - \langle \delta_{R}\kcmb(\theta_{\alpha}) \rangle,
\label{eq:measure_nk}
\end{align}
where
\begin{align}
\langle & \delta_{g}\kcmb(\theta_{\alpha}) \rangle_{0} = \notag \\ &\frac{1}{N^{\delta_{g} \kcmb}_{\theta_{\alpha}}} \sum_{i = 1}^{N_ g} \sum_{j= 1}^{N_{\rm pix}}
\eta^{\delta_{g}}_i \eta^{\kcmb}_j \kappa_{{\rm CMB},j}\Theta_{\alpha}(|\boldsymbol{\hat{\theta}}^i - \boldsymbol{\hat{\theta}}^j|)
\label{eq:nk1}
\end{align}
and
\begin{align}
\langle  &\delta_{R} \kcmb(\theta_{\alpha}) \rangle = \notag \\ &\frac{1}{N^{R\kcmb}_{\theta_{\alpha}}} \sum_{i = 1}^{N_{\rm rand}} \sum_{j= 1}^{N_{\rm pix}}
\eta^{\delta_{R}}_i \eta^{\kcmb}_j \kappa_{{\rm CMB},j}\Theta_{\alpha}(|\boldsymbol{\hat{\theta}}^i - \boldsymbol{\hat{\theta}}^j|),
\end{align}
where the sum in $i$ is over all galaxies and the sum in $j$ is over all pixels in the CMB convergence map; $N^{\delta_{g} \kcmb}_{\theta_{\alpha}}$ ($N^{\delta_{R} \kcmb}_{\theta_{\alpha}}$) is the number of galaxy-$\kcmb{}$ pixel (random-$\kcmb{}$ pixel) pairs that fall within the angular bin $\theta_{\alpha}$; $\eta^{\delta_g}$, $\eta^{\delta_R}$ and $\eta^{\kcmb}$ are the weights associated with the galaxies, the randoms and the $\kcmb$ pixels. The random catalog is used to sample the selection function of the lens galaxies, and has a number density much higher than the galaxies. $\boldsymbol{\hat{\theta}}^{i}$ ($\boldsymbol{\hat{\theta}}^{j}$) is the angular position of galaxy $i$ (pixel $j$), and $\Theta_{\alpha}$ is an indicator function that is 1 if $|\boldsymbol{\hat{\theta}}^i - \boldsymbol{\hat{\theta}}^j|$ falls in the angular bin $\theta_{\alpha}$ and 0 otherwise.

\begin{figure*}
\begin{center}
\includegraphics[width=0.99\linewidth]{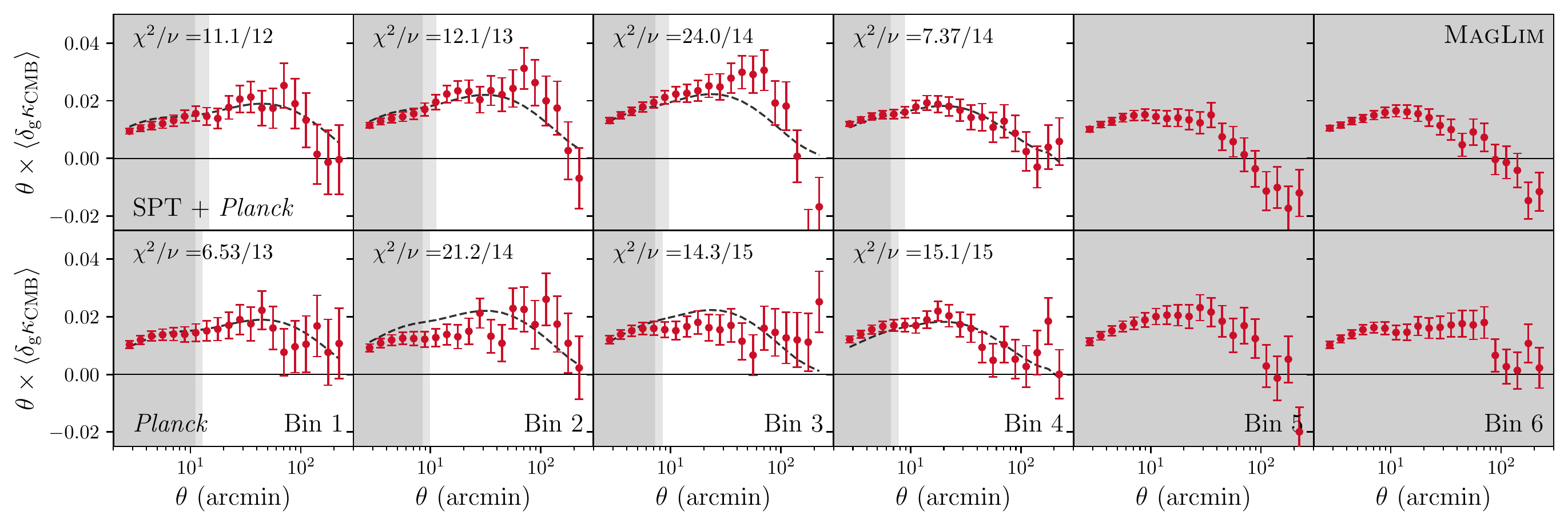}
\includegraphics[width=0.8\linewidth]{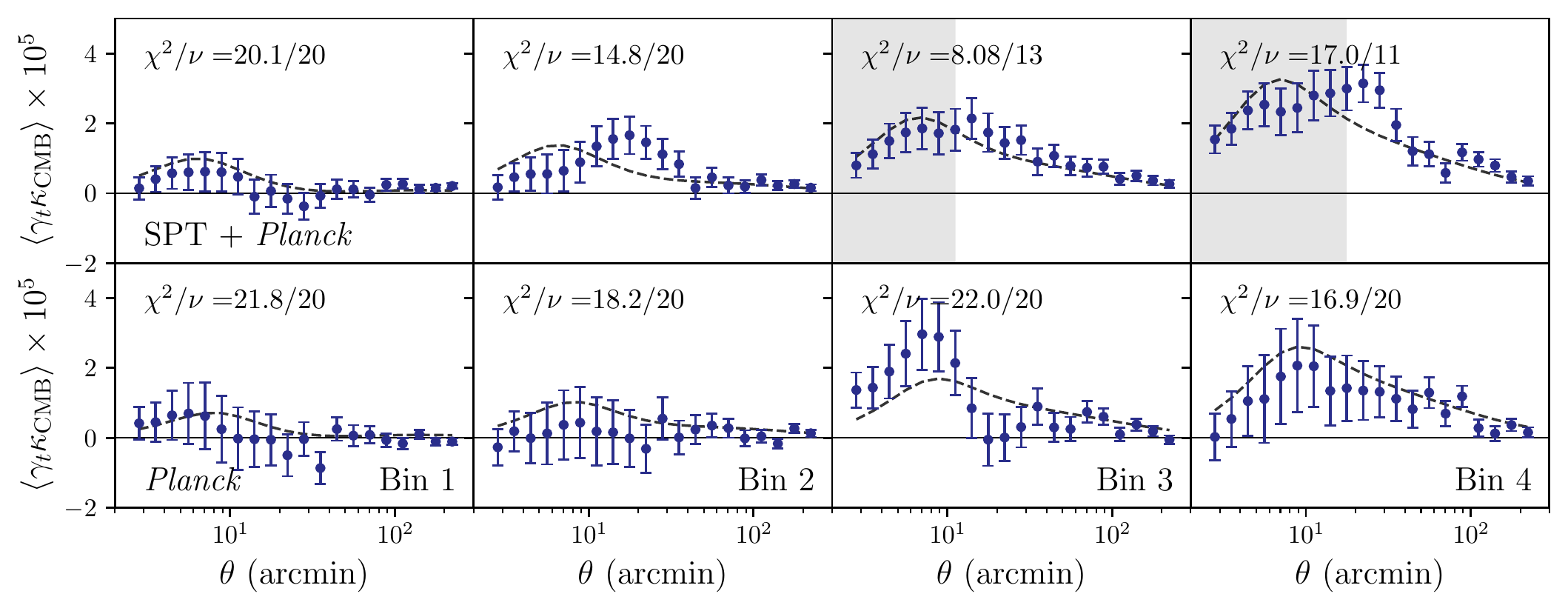}
\caption{Measurement of the \maglim{} galaxy density-CMB lensing correlation (top) and galaxy shear-CMB lensing correlation (bottom). For each set of measurements, the upper row shows measurement with the SPT+{\it Planck} CMB lensing map and the lower row shows measurement with the {\it Planck} CMB lensing map. The shapes and amplitudes are different due to the difference in the $L$ cut and smoothing of the CMB lensing map. The light (dark) shaded regions in the \nk{} panels indicate the data points removed when assuming linear (nonlinear) galaxy bias, while the shaded regions in the \gk{} panels show the data points removed in all cases (only two bins require scale cuts). The dashed dark grey line shows the best-fit fiducial model for the fiducial lens sample, while the $\chi^2$ per degree of freedom ($\nu$) evaluated at the best-fit model with scale cuts for linear galaxy bias model is shown in the upper left corner of each panel.} 
\label{fig:davtavec_maglim}
\end{center}
\end{figure*}

Our estimator for the galaxy shear-CMB lensing correlation (Equation~\ref{eq:wgk}) is
\begin{align}
\langle & \gamma_{\rm t} \kcmb(\theta_{\alpha}) \rangle= \notag \\
&\frac{\sum_{i=1}^{N_{\rm gal}} \sum_{j=1}^{N_{\rm pix}} \eta^{e}_i \eta^{\kcmb}_j  \kappa_{{\rm CMB},j}e_{\rm t}^{ij}\Theta_{\alpha}(|\boldsymbol{\hat{\theta}}^i - \boldsymbol{\hat{\theta}}^j|)}{s (\theta_{\alpha}) \sum \eta^{e}_i \eta^{\kcmb}_j },
\label{eq:measurement}
\end{align}
where $e_{\rm t}^{ij}$ is the component of the corrected ellipticity oriented orthogonally to the line connecting pixel $j$ and the source galaxy.  The $\kcmb$ value in the pixel is $\kcmb^j$ and  $\eta^{e}_i$ and $\eta^{\kcmb}_j$ are the weights associated with the source galaxy and the $\kcmb{}$ pixel, respectively. $s (\theta_{\alpha})$ is the $\metacal$ response. We find that $s (\theta)$ is approximately constant over the angular scales of interest, but different for each redshift bin. 
We carry out these measurements using the $\textsc{TreeCorr}$ package\footnote{\url{https://github.com/rmjarvis/TreeCorr}} \citep{Jarvis2004} in the angular range $2.5'<\theta<250.0'$. Note that Equation~\ref{eq:measurement} does not require subtracting a random component as in  Equation~\ref{eq:measure_nk} since unlike a density field, the mask geometry cannot generate an artificial signal in a shear field.  

\begin{table}
\centering
 \begin{tabular}{l|c|c|c}
\hline
 Scale cuts & None & Linear bias  & Nonlinear \\
 & & & bias \\
 \hline
SPT+{\it Planck} & & & \\
\nk{} \maglim{} & 26.8 & 14.5  & 17.3 \\ 
\textcolor{gray}{\nk{} \maglim{} 6 bin}& \textcolor{gray}{30.2} & \textcolor{gray}{17.4} &  \textcolor{gray}{20.0}\\ 
\textcolor{gray}{\nk{} \redmagic{}} & \textcolor{gray}{23.7} & \textcolor{gray}{14.2} & \textcolor{gray}{15.7} \\ 
\gk{} \maglim{} & 15.0 & 13.4 & 13.4\\
\hline
{\it Planck} & && \\
\nk{} \maglim{}& 17.9 & 13.1 & 13.8\\ 
\textcolor{gray}{\nk{} \maglim{} 6 bin} & \textcolor{gray}{20.5} & \textcolor{gray}{15.9} & \textcolor{gray}{16.8} \\ 
\textcolor{gray}{\nk{} \redmagic{}} & \textcolor{gray}{17.0} & \textcolor{gray}{12.5} & \textcolor{gray}{12.8} \\ 
\gk{} \maglim{} & 10.4 & 10.4 & 10.4 \\
\hline
Combined &  &  & \\
\nk{} \maglim{} & 32.2 & 19.6& 22.2 \\ 
\gk{} \maglim{} & 18.2 & 16.9& 16.9\\
\nkgk{} \maglim{} & 34.8 &23.9 & 25.7 \\ 
\hline
\end{tabular}
\caption{Signal-to-noise for the different parts of the \nkgk{} data vector when different scale cuts are applied. Rows involving the two high-redshift \maglim{} bins and the \redmagic{} sample are shown in grey to indicate that they are not part of the fiducial analysis.} 
\label{table:s2n}
\end{table}

The measured \maglim{} \nk{} and \gk{} correlation functions are shown in Figure~\ref{fig:davtavec_maglim}. The \nk{} measurements using the \redmagic{} sample are shown in Appendix~\ref{sec:redmagic}. The signal-to-noise (S/N) of the different measurements are listed in Table~\ref{table:s2n}. Here, signal-to-noise is calculated via
\begin{equation}
    {\rm S/N}\equiv \sqrt{\sum^N_{ij} d_i^T \mathbf{C}^{-1}_{ij} d_j}, 
\end{equation}
where $d$ is the data vector of interest and $\mathbf{C}$ is the covariance matrix. The final signal-to-noise of the fiducial \nkgk{} data vector after the linear bias scale cuts is 23.9, about two times larger than in the Y1 study \citep{y15x2} -- the main improvement, in addition to the increased sky area, comes from extending our analysis to smaller scales, enabled by the tSZ-cleaned CMB lensing map. The tSZ signal is correlated with large-scale structure, and can propagate into a bias in the estimated $\kappa_{\rm CMB}$ if not mitigated. In the DES Y1 analysis presented in \cite{y15x2}, tSZ cleaning was not implemented at the $\kappa_{\rm CMB}$ map level, necessitating removal of small-scale CMB lensing correlation measurements from the model fits. This problem was particularly severe for \gk{}. Comparing results for the SPT+{\it Planck} and {\it Planck} patches in Table~\ref{table:s2n}, the SPT+{\it Planck} area dominates the signal-to-noise before scale cuts in all the probes, even with a smaller sky area. This is due to the lower noise level of the SPT maps. However, since the higher signal-to-noise necessitates a more stringent scale cut, the resulting signal-to-noise after scale cuts is only slightly higher for the SPT+{\it Planck} patch. Finally, comparing \nk{} and \gk{}, even though \nk{} starts with $\sim$75\% more signal-to-noise before scale cuts compared to \gk{}, the scale cuts remove significantly more signal in \nk{} compared to \gk{}. This is due to limits in our ability to model nonlinear galaxy bias on small scales -- indeed we see that the signal-to-noise in \nk{} increases by 13\% when switching from linear to nonlinear galaxy bias model.   Overall, these signal-to-noise levels are consistent with the forecasts in \citetalias{y3-nkgkmethods}.

\section{Blinding and unblinding}
\label{sec:blinding}

Following \cite{y3-3x2ptkp}, we adopt a strict, multi-level blinding procedure in our analysis designed to minimize the impact of experimenter bias. The first level of blinding occurs at the shear catalog level, where all shears are multiplied by a secret factor \cite{y3-shapecatalog}. The second level of blinding occurs at the two-point function level, where we follow the procedure outlined in \cite{y3-blinding} and shift the data vectors by an unknown amount while maintaining the degeneracy between the different parts of the data vector under the same cosmology. The main analyses in this paper were conducted after the unblinding of the shear catalog, so the most relevant blinding step is the data vector blinding. Below we outline the list of tests that were used to determine whether our measurement is sufficiently robust to unblind:
\begin{itemize}
    \item Pass all tests described in Appendix~\ref{sec:sys_test}, which indicate no outstanding systematic contamination in the data vectors. These tests include: (1) check for spurious correlation of our signal with survey property maps, (2) check the cross-shear component of \gk{}, (3) check the impact of weights used for the lens galaxies, (4) check the effect of the point-source mask in the CMB lensing map on our measurements,  and (5) check that cross-correlating an external large-scale structure tracer (the cosmic infrared background in this case) with different versions of our CMB lensing maps yields consistent results. 
    \item With unblinded chains, use the posterior predictive distribution (PPD) method developed in \citep{y3-inttensions} to evaluate the consistency between the two subsets of the data vectors that use different CMB lensing maps (i.e. the SPT+{\it Planck} patch and the {\it Planck} patch). The $p$-value should be larger than 0.01.
    \item With unblinded chains, verify that the goodness-of-fit of the data with respect to the fiducial model has a $p$-value larger than 0.01 according to the same PPD framework.
\end{itemize}
Except for the first step, all the above are applied to the \nkgk{} data vectors with the fiducial analysis choices ($\Lambda$CDM cosmology and linear galaxy bias scale cuts), for the first four bins of the \maglim{} lens sample.   

\section{Parameter constraints from cross-correlations of DES with CMB lensing}
\label{sec:results}

Following the steps outlined in the previous section, we found (1) no evidence for significant systematic biases in our measurements, as shown in Appendix~\ref{sec:sys_test}, (2) we obtain a $p$-value greater than 0.01 
when comparing the \nkgk{} constraints from the {\it Planck} region
to constraints from the SPT+{\it Planck} region, and (3) the goodness-of-fit test of the fiducial \nkgk{} unblinded chain has a $p$-value greater than 0.01. In the following, we will quote the precise $p$-values obtained from these tests using the updated covariance matrix.

With all the unblinding tests passed, we froze all analysis choices and unblinded our cosmological constraints. We then updated the covariance matrix  to match the best-fit parameters from the cosmological analysis.\footnote{This procedure is the same as in \cite{y3-3x2ptkp}. Since we can not know the cosmological and nuisance parameters exactly before running the full inference, a set of fiducial parameters were used to generate the first-pass of the covariance that was used for all blinded chains. After unblinding, we update the parameters to values closer to the best-fit parameters from the data. After confirming that the \fivetwo{} best-fit constraints \citetalias{y3-5x2kp} are consistent with the \threetwo{} best-fit constraints, we chose to use the \threetwo{} best-fit parameters for evaluating the covariance matrix, as this makes our modeling choices more consistent with that of \cite{y3-3x2ptkp}.} The results we present below use the updated covariance matrix. The main constraints on cosmological parameters are summarized in Table~\ref{table:cosmo}.

\subsection{Cosmological constraints from cross-correlations}

\begin{figure} 
\begin{center}
\includegraphics[width=0.99\linewidth]{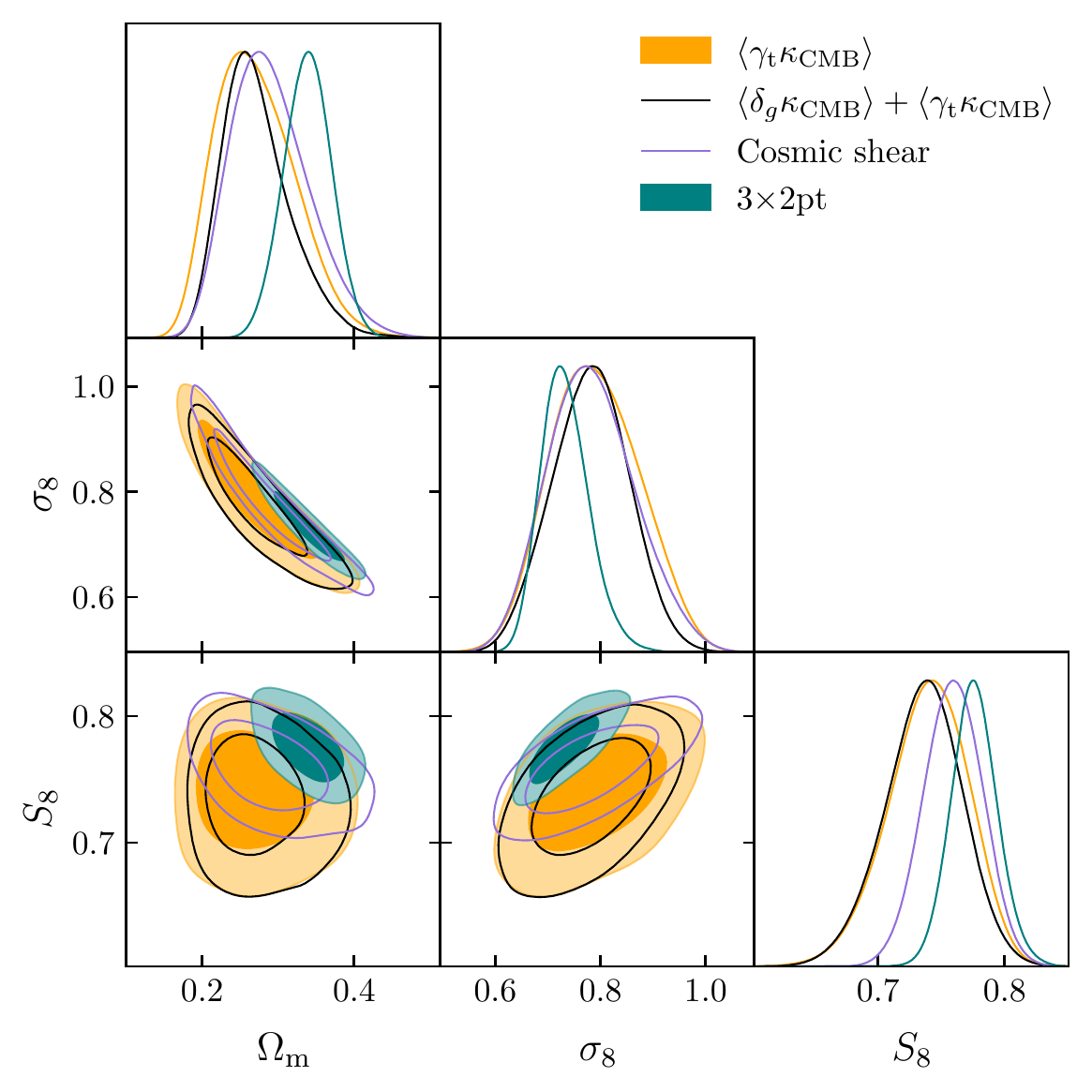}
\caption{Constraints on cosmological parameters $\Omega_{\rm m}$, $\sigma_{8}$, and $S_{8}$ from \nkgk{} using the \maglim{} sample. We also show the corresponding constraints from \gk{}-only, cosmic shear and \threetwo{} for comparison. }
\label{fig:main}
\end{center}
\end{figure}

In Figure~\ref{fig:main} we show constraints from \nkgk{} using the first 4 bins of the \maglim{} sample. For comparison, we also show constraints from \gk{}-only, cosmic shear (from \cite{y3-cosmicshear1,y3-cosmicshear2}), and \threetwo{} (from \cite{y3-3x2ptkp}).

We find that our analysis of \nkgk{} gives the following constraints:
\begin{align}
    \Omega_{\rm m} &= 0.272^{+0.032}_{-0.052}; \notag \\
    \sigma_{8} &= 0.781^{+0.073}_{-0.073}; \notag \\
    S_{8} &= 0.736^{+0.032}_{-0.028}.  \notag
\end{align}

As can be seen from Figure~\ref{fig:main} and expected from \citetalias{y3-nkgkmethods}, the constraints are dominated by \gk{}, with \nk{} slightly improving the $\Omega_{\rm m}$ constraints.  While \nk{} by itself does not tightly constrain cosmology because of the degeneracy with galaxy bias, the shape information in \nk{} provides additional information on $\Omega_m$ when combined with \gk{}.
 
Figure~\ref{fig:main} also shows constraints from DES-only probes, including cosmic shear and \threetwo{}.  We find that the constraints on $S_8$ from \nkgk{} are comparable to those from cosmic shear and \threetwo{}, and in reasonable agreement. The uncertainties of the \nkgk{} constraints on $S_{8}$ are roughly 30\% (70\%) larger than that of cosmic shear (\threetwo{}). We will perform a complete assessment of consistency between these probes in \citetalias{y3-5x2kp}.   We can also see that the degeneracy direction of the \nkgk{} constraints are slightly different from \threetwo{}, which will help in breaking degeneracies when combined. 

\begin{table*}
\centering
 \begin{tabular}{lcccc}
\hline
Dataset  & $\sigma_8$ & $\Omega_m$ & $S_8$ & PPD $p$-value \\
\hline
\gk{} \maglim{} & $0.790^{+0.080}_{-0.092}$ & 
$0.270^{+0.043}_{-0.061}$ & 
$0.740^{+0.034}_{-0.029}$ & 0.72\\
\nkgk{} \maglim{} 4 bin linear galaxy bias & $0.781^{+0.073}_{-0.073}$ & 
$0.272^{+0.032}_{-0.052}$ & 
$0.736^{+0.032}_{-0.028}$ & 0.50\\
\nkgk{} \maglim{} 4 bin nonlinear galaxy bias & $0.820^{+0.079}_{-0.067}$ & 
$0.245^{+0.026}_{-0.044}$ & 
$0.734^{+0.035}_{-0.028}$ & 0.51\\
\hline 
\textcolor{gray}{\nkgk{} \maglim{} 6 bin linear galaxy bias} & \textcolor{gray}{$0.755^{+0.071}_{-0.071}$} & \textcolor{gray}{$0.288^{+0.037}_{-0.053}$} & \textcolor{gray}{$0.732^{+0.032}_{-0.029}$} & \textcolor{gray}{0.45}\\
\textcolor{gray}{\nkgk{} \maglim{} 6 bin nonlinear galaxy bias} & \textcolor{gray}{$0.769^{+0.071}_{-0.071}$} & \textcolor{gray}{$0.273^{+0.034}_{-0.047}$} & \textcolor{gray}{$0.727^{+0.035}_{-0.028}$}& \textcolor{gray}{0.45}\\
\textcolor{gray}{\nkgk{} \redmagic{} linear galaxy bias} & \textcolor{gray}{$0.793^{+0.072}_{-0.083}$} & \textcolor{gray}{$0.266^{+0.036}_{-0.050}$} & \textcolor{gray}{$0.738^{+0.034}_{-0.030}$} & \textcolor{gray}{0.39}\\
\textcolor{gray}{\nkgk{} \redmagic{} nonlinear galaxy bias} & \textcolor{gray}{$0.794^{+0.069}_{-0.069}$} & \textcolor{gray}{$0.253^{+0.030}_{-0.046}$} & \textcolor{gray}{$0.723^{+0.033}_{-0.030}$} & \textcolor{gray}{0.41}\\
\hline
\end{tabular}
\caption{$\Lambda$CDM constraints on $\Omega_{\rm m}$, $\sigma_{8}$ and $S_{8}$ using \nkgk{} and different lens samples. We show the constraints using both linear and nonlinear galaxy bias. The last column shows the $p$-value corresponding to the goodness of fit for the chain. The parts shown in grey indicate that they are not part of the fiducial samples.} 
\label{table:cosmo}
\end{table*}

\begin{figure} 
\begin{center}
\includegraphics[width=0.99\linewidth]{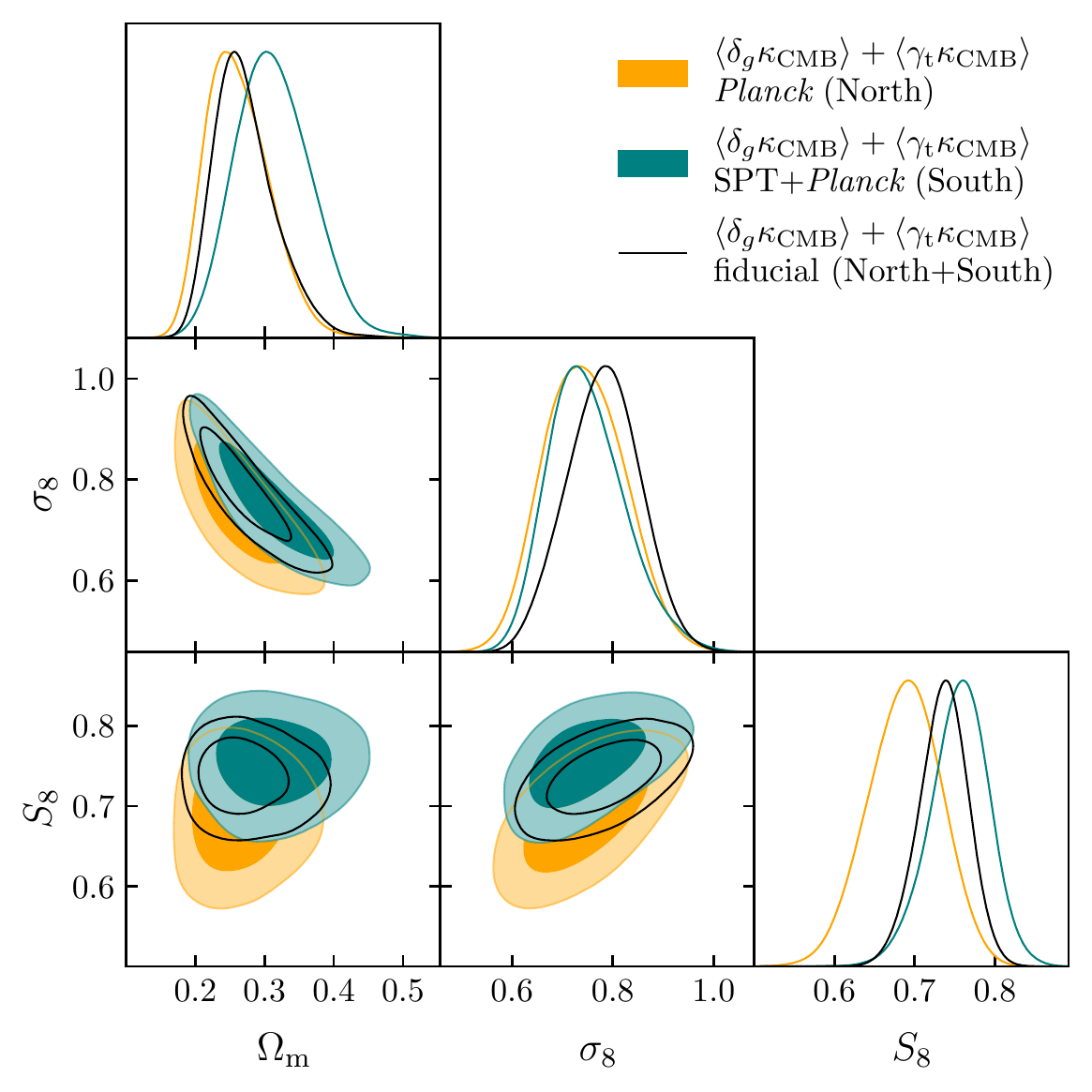}
\caption{Constraints on cosmological parameters $\Omega_{\rm m}$, $\sigma_{8}$, and $S_{8}$ using the \nkgk{} probes. We also show the constraints only using the SPT+{\it Planck} area and only using the {\it Planck} area.}
\label{fig:nkgk_planck_spt}
\end{center}
\end{figure}

We consider constraints from the SPT+{\it Planck} and {\it Planck} patches separately in Figure~\ref{fig:nkgk_planck_spt}. As discussed earlier in Section~\ref{sec:blinding}, the consistency of these two patches was part of the unblinding criteria, thus these two constraints are consistent under the PPD metric. We find a $p$-value of 0.37 (0.33) when comparing the {\it Planck} (SPT+{\it Planck}) results to constraints from SPT+{\it Planck} ({\it Planck}).  We also observe that the constraints are somewhat tighter in the SPT+{\it Planck} patch in $S_8$, consistent with the slightly larger signal-to-noise (see Table~\ref{table:s2n}). We note however, that the signal-to-noise before scale cuts of the SPT+{\it Planck} patch is significantly larger than the {\it Planck} patch due to the lower noise and smaller beam size of the SPT lensing map (for \nk{}: 26.8 vs. 17.9; for \gk{}: 15.0 vs. 10.4), though most of the signal-to-noise is on the small scales which we had to remove due to uncertainties in the theoretical modeling. This highlights the importance of improving the small-scale modeling in future work.

\begin{figure} 
\begin{center}
\includegraphics[width=0.99\linewidth]{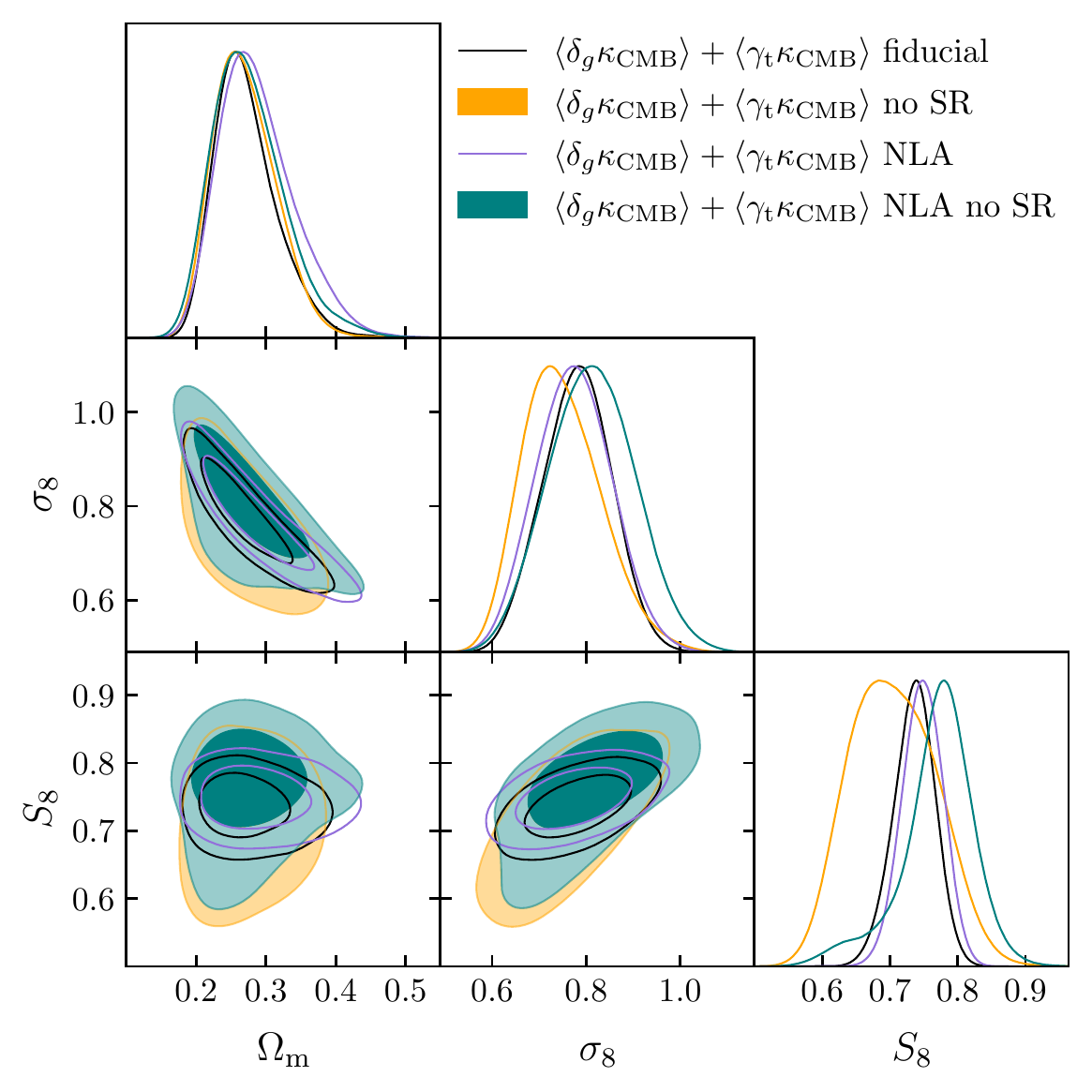}
\caption{Constraints on cosmological parameters $\Omega_{\rm m}$,$\sigma_{8}$, and $S_{8}$ using the \nkgk{} probes with and without including the lensing ratio (SR) likelihood, and when assuming the NLA IA model instead of our fiducial IA model TATT.}
\label{fig:IA}
\end{center}
\end{figure}

\begin{figure*} 
\begin{center}
\includegraphics[width=0.85\linewidth]{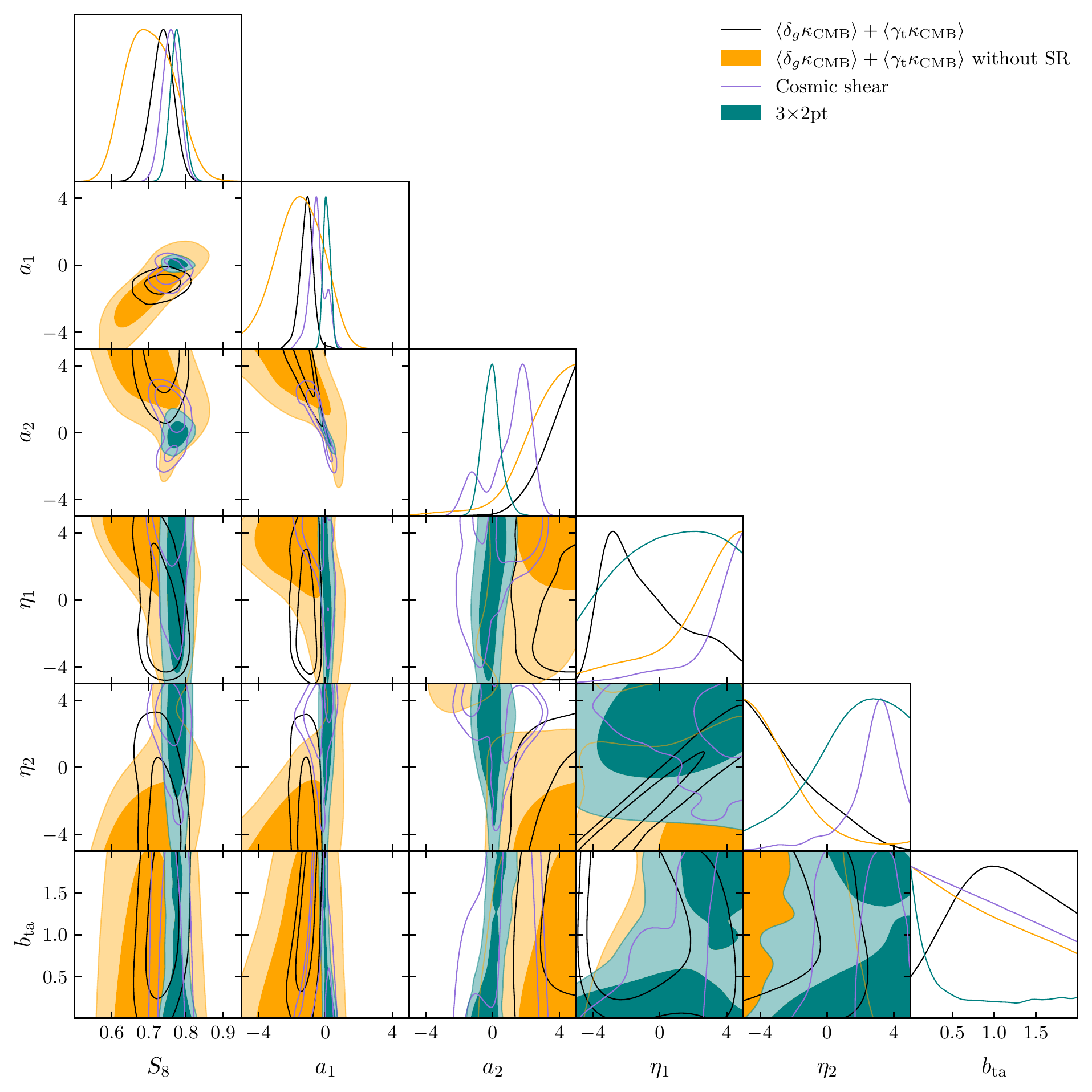}
\caption{Constraints on $S_8$ and the IA parameters from our fiducial \nkgk{} results, cosmic shear and \threetwo{}. We also include the \nkgk{} constraints without the lensing ratio (SR) likelihood for comparison.}
\label{fig:IA2}
\end{center}
\end{figure*}

\begin{figure} 
\begin{center}
\includegraphics[width=0.99\linewidth]{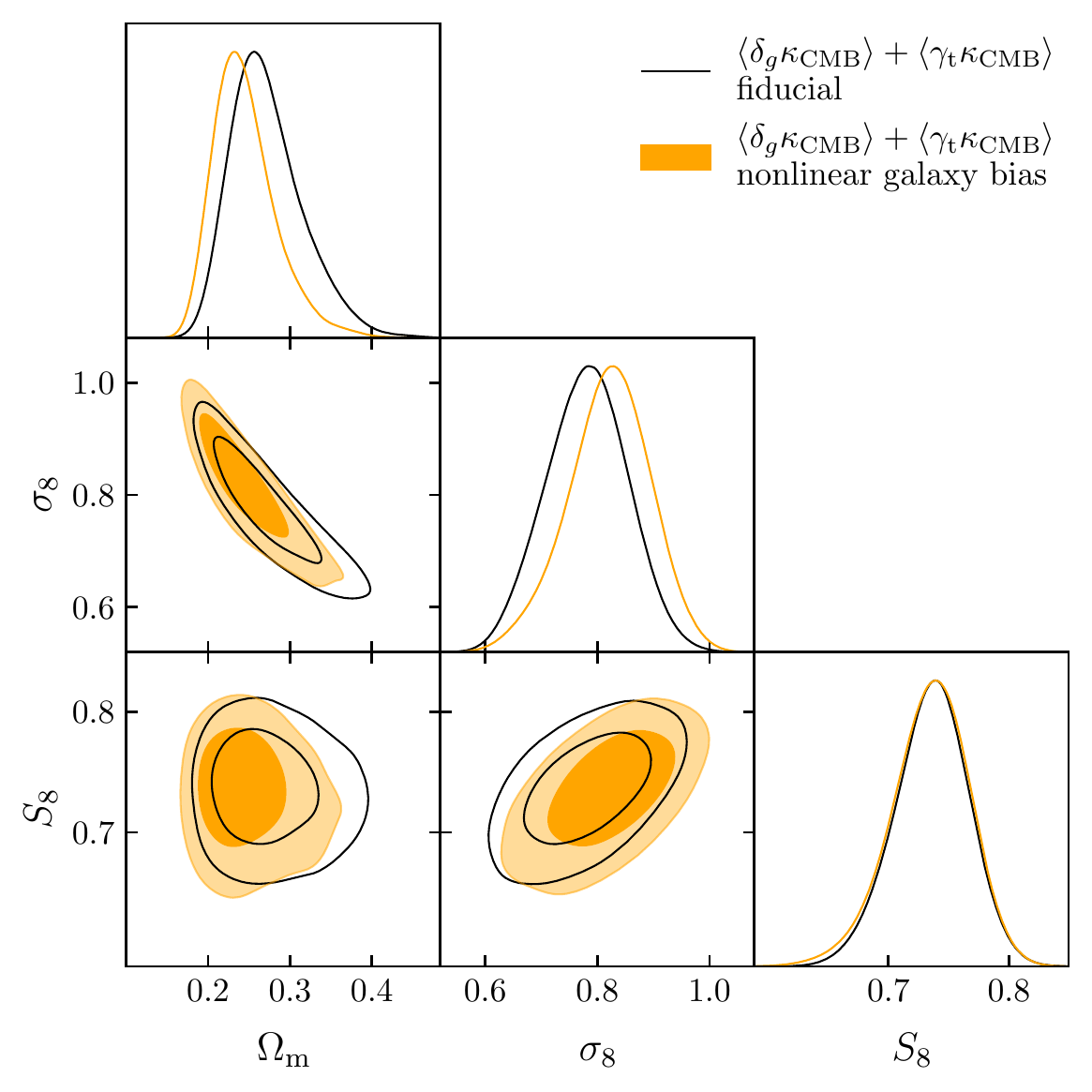}
\caption{Fiducial \nkgk{} constraints on cosmological parameters $\Omega_{\rm m}$, $\sigma_{8}$, and $S_{8}$ using linear and nonlinear galaxy bias models.}
\label{fig:gal_bias}
\end{center}
\end{figure}

\subsection{Lensing ratio and IA modeling}
\label{sec:SR_IA}

As discussed in Section~\ref{sec:model}, we have included the lensing ratio likelihood in all our constraints. As was investigated in detail in \cite{y3-shearratio}, the inclusion of the lensing ratio information mainly constrains the IA parameters and source galaxy redshift biases. The TATT IA model adopted here is a general and flexible model that allows for a large range of possible IA contributions. As such, it is expected that including the lensing ratio could have a fairly large impact for data vectors that are not already constraining the IA parameters well. We now examine the effect of the lensing ratio on our fiducial \nkgk{} constraints by first removing the lensing ratio prior in our fiducial result, and then doing the same comparison with a different, more restrictive IA model, the NLA model (see Section~\ref{sec:model}). These results are shown in Figure~\ref{fig:IA}.

We make several observations from Figure~\ref{fig:IA}. First, the lensing ratio significantly tightens the constraints in the $S_8$ direction (roughly a factor of 2), as expected from \citetalias{y3-nkgkmethods}. Second, without the lensing ratio, different IA models result in different $S_8$ constraints, with TATT resulting in $\sim 40\%$ larger uncertainties than NLA. This is expected given that TATT is a more general model with three more free parameters to marginalize over compared to NLA. That being said, the constraints are still fully consistent when using the different IA models. Third, when lensing ratio is included, there is very little difference in the constraints between the two different IA models. This suggests that the IA constraints coming from the lensing ratio are sufficient to make the final constraints insensitive to the particular IA model of choice. 

Finally, it is interesting to look at the constraints on the IA parameters for our fiducial \nkgk{} analysis with and without the lensing ratio. We show this in Figure~\ref{fig:IA2}, and compare them with constraints from cosmic shear \cite{y3-cosmicshear1,y3-cosmicshear2} and \threetwo{} \cite{y3-3x2ptkp}. We find two noticeable degeneracies in these parameters: 
\begin{itemize}
\item The lensing ratio restricts the $a_1-a_2$ parameter space to a narrow band. This is seen  in the cosmic shear and \threetwo{} results, as well as the \nkgk{} results, although \nkgk{} prefers somewhat higher $a_2$ values.
\item There is a noticeable $\eta_1-\eta_2$ degeneracy that shows up uniquely in \nkgk{} and not in the other probes in the plot. We note that this degeneracy is likely sourced by the lensing ratio likelihood, which on its own is degenerate in the $\eta_1-\eta_2$ plane. This is consistent with what we have seen in the simulations in \citetalias{y3-nkgkmethods}. The fact that it appears more prominent in \nkgk{} than in the other probes is partly related to the fact that $a_1$ and $a_2$ are constrained to be further away from zero in the case of \nkgk{}, allowing $\eta_1$ and $\eta_2$ (the redshift evolution of the terms associated with $a_1$ and $a_2$) to be constrained better. Another relevant factor is that \nkgk{} probes slightly larger redshift ranges than cosmic shear and \threetwo{} due to the CMB lensing kernal, which allows for a longer redshift lever arm to constrain $\eta_1$ and $\eta_2$, resulting in qualitatively different behaviors in the $\eta_1-\eta_2$ parameter space.
\end{itemize} 

\begin{figure} 
\begin{center}
\includegraphics[width=0.99\linewidth]{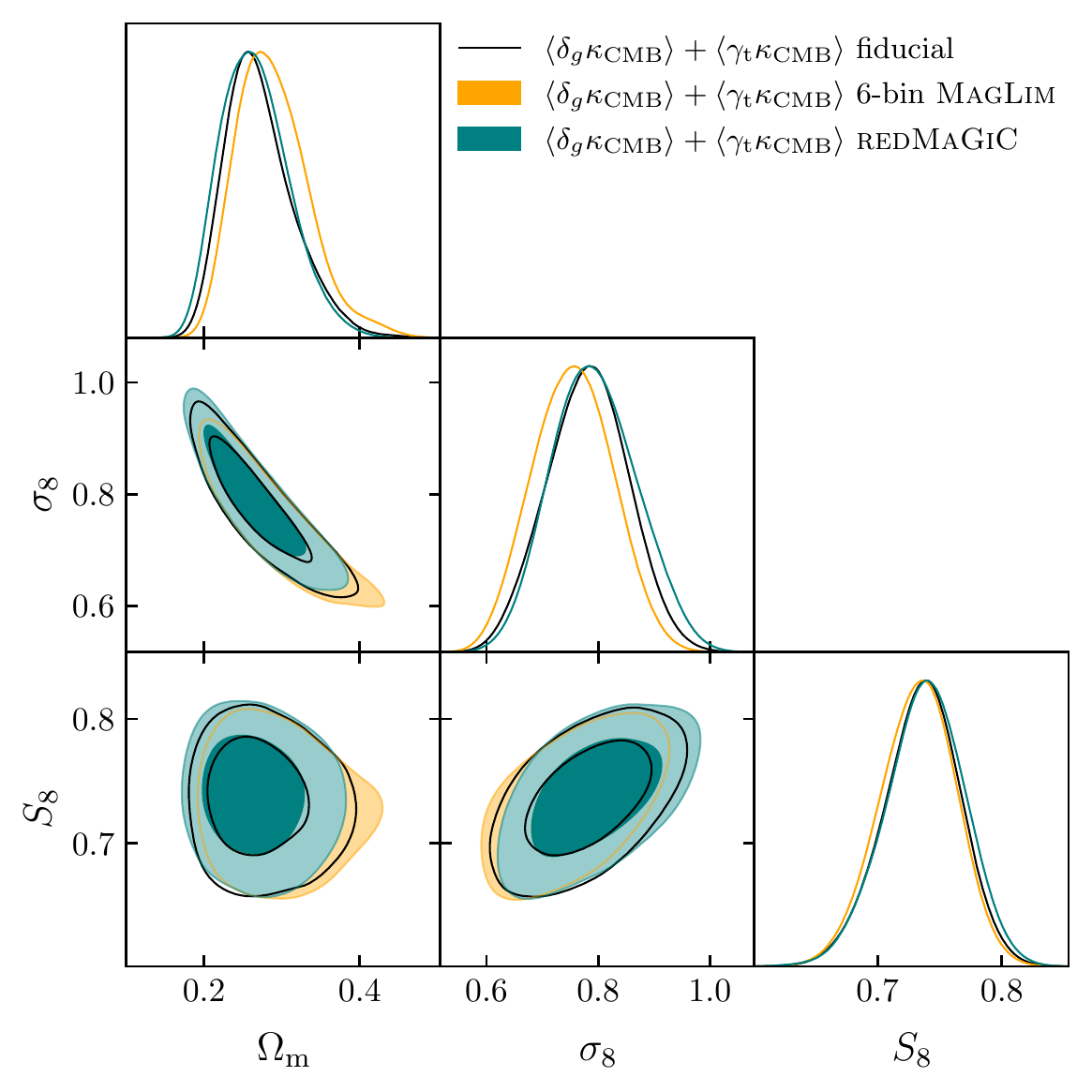}
\caption{Fiducial constraints on cosmological parameters $\Omega_{\rm m}$, $\sigma_{8}$, and $S_{8}$ using the \nkgk{} probes compared with using the \redmagic{} lens sample instead of the \maglim{} lens sample.}
\label{fig:maglimvsredmagic}
\end{center}
\end{figure}

\begin{figure*} 
\begin{center}
\includegraphics[width=0.99\linewidth]{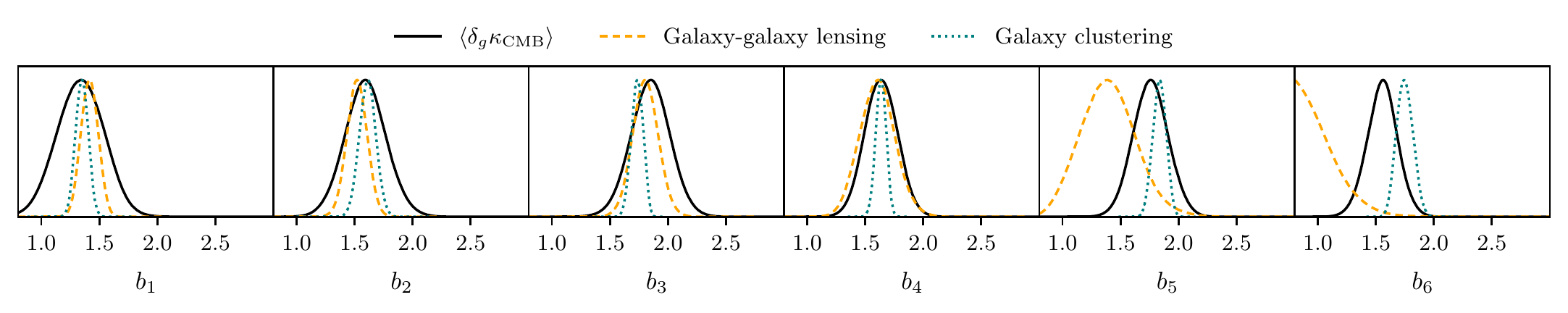}
\includegraphics[width=0.85\linewidth]{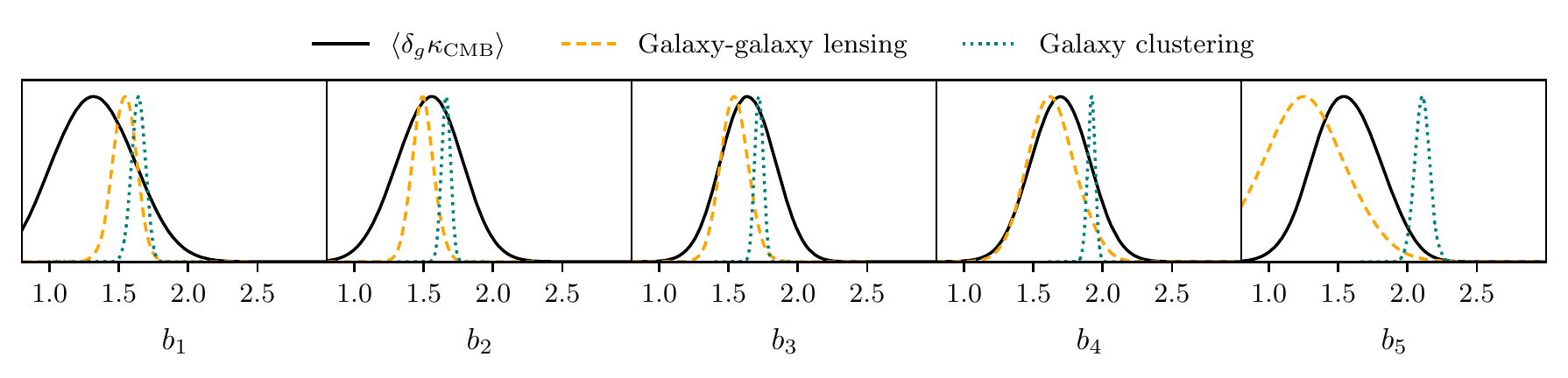}
\caption{With fixed cosmological parameters, the inferred galaxy bias from \nk{}, galaxy-galaxy lensing and galaxy clustering, for the \maglim{} sample (top) and the \redmagic{} sample (bottom).}
\label{fig:Xlens}
\end{center}
\end{figure*}

\subsection{Nonlinear galaxy bias}

As discussed in Section~\ref{sec:model}, we test a nonlinear galaxy bias model in addition to our baseline linear galaxy bias analysis. With a nonlinear galaxy bias model we are able to use somewhat smaller scales and utilize more signal in the data (see Table~\ref{table:s2n}). In Figure~\ref{fig:gal_bias} we show the cosmological constraints of our fiducial \nkgk{} data vector with the nonlinear galaxy bias model. We find that the constraints between the two different galaxy bias models are consistent. There is a small improvement in the $\Omega_{\rm m}$ direction, which is not surprising given that nonlinear bias impacts \nk{}, and \nk{} improves the $\Omega_{\rm m}$ constraints relative to \gk{} alone. The overall improvement is nevertheless not very significant, as \gk{} is dominating the constraints.

\subsection{Comparison with alternative lens choices}
\label{sec:Xlens}

We have defined our fiducial lens sample to be the first four bins of the \maglim{} sample. This choice is informed by the \threetwo{} analysis in \cite{y3-3x2ptkp}, where alternative lens samples were also tested but were deemed to be potentially contaminated by systematic effects and therefore not used in the final cosmology analysis. Here, we examine the \nkgk{} constraints using the two alternative choices for lenses: (1) including the two high-redshift bins in \maglim{} to form a 6-bin \maglim{} sample, and (2) the \redmagic{} lens sample. As we have emphasized throughout the paper, since the galaxy-CMB lensing cross-correlation is in principle less sensitive to some of the systematic effects, these tests could potentially shed light on the issues seen in \cite{y3-3x2ptkp}. We only examine the \nkgk{} constraints here, but will carry out a more extensive investigation in combination with the \threetwo{} probes in \citetalias{y3-5x2kp}.

In Figure~\ref{fig:maglimvsredmagic} we show constraints from \nkgk{} using the three different lens samples: 4-bin \maglim{} (fiducial), 6-bin \maglim{} and \redmagic{}. The best-fit parameters as well as the goodness-of-fit are listed in Table~\ref{table:cosmo}. Broadly, all three constraints appear to be very consistent with each other. This is not surprising given that the constraining power is dominated by \gk{} as we discussed earlier.  In \cite{y3-3x2ptkp} it was shown that for the \threetwo{} analysis, both the 6-bin \maglim{} and the \redmagic{} samples give goodness-of-fits that fail our criteria, while for \nkgk{} all three samples give acceptable goodness-of-fits values as seen in Table~\ref{table:cosmo}. This could imply that the systematic effects that contaminated the other correlation functions in \threetwo{} are not affecting the \nkgk{} results strongly. Compared to the fiducial constraints, the constraining power of the 6-bin \maglim{} sample is slightly higher in the $\Omega_{\rm m}$ direction due to the added signal-to-noise from the high-redshift bins, while the constraining power of the \redmagic{} sample is slightly lower in both $\Omega_{\rm m}$ and $S_8$.

The DES Y3 \threetwo{} analyses found that the poor fits for the alternative lens samples can be explained by inconsistent galaxy bias between galaxy-galaxy lensing $\langle \delta_{g} \gamma_{\rm t} \rangle$ and galaxy clustering $\langle \delta_{g}\delta_{g}  \rangle$. That is, when allowing the galaxy bias to be different in galaxy-galaxy lensing and galaxy clustering, the goodness-of-fit improves significantly. Operationally, this is achieved in \cite{y3-3x2ptkp} by adding a free parameter, $X_{\rm lens}$, defined such that

\begin{equation}
X_{\rm lens}^i = b^i_{\langle \delta_g \gamma_{\rm t}\rangle} / b^i_{\langle \delta_g \delta_g \rangle},
\end{equation}
where $b^i_{\langle \delta_g \gamma_{\rm t}\rangle}$ ($b^i_{\langle \delta_g \delta_g \rangle}$) is the linear galaxy bias parameter for $\langle \delta_{g}\gamma_{\rm t}\rangle$ ($\langle \delta_{g}\delta_{g}\rangle$) in lens galaxy redshift bin $i$. 
% \begin{align}
%     \langle \delta_{g}\delta_{g}\rangle & = b^2 \xi_{\rm mm}, \notag \\
%     \langle \delta_{g}\gamma_{\rm t}\rangle & =  X_{\rm lens} b \xi_{\rm mm},
% \end{align}
% where $\xi_{\rm mm}$ is the two-point matter-matter correlation function. 
$X_{\rm lens}$ is expected to equal 1 in the case of no significant systematic effects. In \cite{y3-3x2ptkp} it was found that $X_{\rm lens} \neq 1$ for the two high-redshift bins in the \maglim{} sample and for all bins in the \redmagic{} sample, though there was not enough information to determine whether the systematic effect was in $\langle \delta_{g}\gamma_{\rm t}\rangle$ or $\langle \delta_{g}\delta_{g}\rangle$.

Our CMB lensing cross-correlation analysis provides an interesting way to explore this systematic effect.  In essence, with fixed cosmology, we can fit for galaxy bias using \nk{} and compare with the galaxy bias derived from $\langle \delta_{g}\gamma_{t}\rangle$ and $\langle \delta_{g}\delta_{g}\rangle$. Our results are shown in Figure~\ref{fig:Xlens}. We find that in general the constraints from \nk{} on galaxy bias are weaker than both galaxy-galaxy lensing and galaxy clustering, this is expected due to the lower signal-to-noise. As such, the \nk{}-inferred galaxy bias values are largely consistent with both galaxy-galaxy lensing and galaxy clustering. There are a few bins, though, where \nk{} does show a preference for the galaxy bias values to agree more with one of the two probes. Noticeably, for the last two \maglim{} bins, \nk{} prefers a galaxy bias value that is closer to that inferred by galaxy clustering. On the other hand, for the highest two \redmagic{} bins, \nk{} prefers galaxy bias values that are closer to galaxy-galaxy lensing. These findings are consistent with the various investigations on $X_{\rm lens}$ described in \cite{y3-2x2ptbiasmodelling} and \cite{y3-2x2ptaltlensresults} and suggest potential issues in the measurements or modeling of galaxy-galaxy lensing in the two high-redshift \maglim{} bins and galaxy clustering in the \redmagic{} sample\footnote{In particular, \cite{y3-2x2ptbiasmodelling} tested an alternative \redmagic{} sample and suggested potential remedies to the systematic effect in \redmagic{} that will be explored in future work.}. However, we caution that these results can be cosmology-dependent, and change slightly if a different cosmology is assumed.

\begin{figure*}
    \centering
    \includegraphics[width=0.87\linewidth]{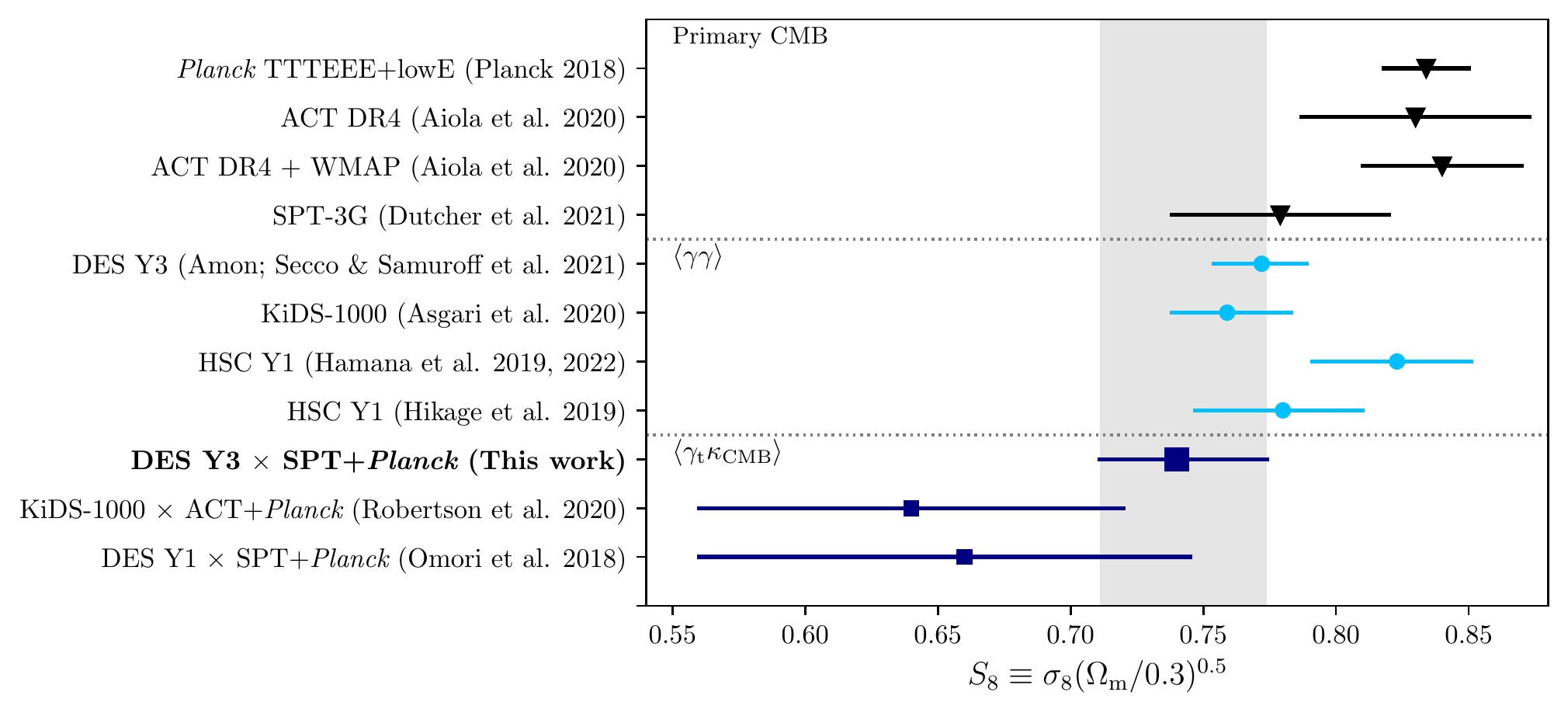}
    \caption{Comparison of late-time measurements of $S_8$ from lensing-only data (cosmic shear $\langle \gamma \gamma \rangle$ and galaxy shear-CMB lensing cross-correlation \gk{}) to the inferred value of $S_8$ from 
    the primary CMB.} 
    \label{fig:S8_compare}
\end{figure*}

\subsection{Implications for $S_8$ tension}

In Figure~\ref{fig:S8_compare}, we compare our constraints on $S_8$ from \gk{} to those from recent measurements of cosmic shear from galaxy surveys (light blue circles) as well as other recent \gk{} constraints (dark blue squares). We show only the constraint from \gk{} (rather than \nkgk{}) since we want to compare only measurements of gravitational lensing. These lensing measurements are not sensitive to the details of galaxy bias, unlike \nk{}. We see that the constraints on $S_8$ obtained from \gk{} in this work (grey band) are for the first time comparable to the state-of-the-art cosmic shear measurements.

Figure~\ref{fig:S8_compare} also shows the inferred value of $S_8$ from the primary CMB (black triangles), as measured by {\it Planck} \citep{Planck:cmblensing}, ACT DR4 \citep{Aiola2020}, combining ACT DR4 and the Wilkinson Microwave Anisotropy Probe \citep[WMAP,][]{Aiola2020}, and SPT-3G \citep{Dutcher2021}. As discussed in several previous works \citep[e.g.][]{Asgari2019,y3-cosmicshear1,y3-cosmicshear2} and can be seen in the Figure, there is a $\sim$2.7$\sigma$ tension\footnote{Here we are quoting the 1D parameter difference in $S_{8}$, or $(S_{8}^{1}-S_{8}^{2})/\sqrt{\sigma^2(S_{8}^1)+\sigma^2(S_{8}^2)}$, where the superscript 1 and 2 refer to the two datasets we are comparing.} between the $S_8$ value inferred from cosmic shear and the {\it Planck} primary CMB constraint -- cosmic shear results prefer a lower $S_8$ value. 
This is intriguing given that it could indicate an inconsistency in the $\Lambda$CDM model. We also see that the other CMB datasets are currently much less constraining, but show some variation, with the lowest $S_8$ value from SPT-3G fairly consistent with all the cosmic shear results.

With this work, we can now meaningfully add \gk{} into this comparison, and as we see in Figure~\ref{fig:S8_compare}, the \gk{} constraints on $S_8$ are also largely below that coming from the primary CMB.  This is potentially exciting, since the \gk{} measurements come from a cross-correlation between two very different surveys, and are therefore expected to be highly robust to systematic errors. Our results therefore lend support to the existence of the $S_8$ tension. In \citetalias{y3-5x2kp} we will perform a more rigorous and complete analysis of the consistency of our constraints here with other datasets.  

\section{Summary}
\label{sec:discussion}

We have presented measurements of two cross-correlations between galaxy surveys and CMB lensing: the galaxy position-CMB lensing correlation (\nk{}), and the galaxy shear-CMB lensing correlation (\gk{}). These measurements are sensitive to the statistics of large-scale structure, and are additionally expected to be very robust to many observational systematics. Our measurements make use of the latest data from the first three years of observations of DES, and a new CMB lensing map constructed explicitly for cross-correlations using SPT and {\it Planck} data. In particular, our fiducial results are from four tomographic bins of the \maglim{} lens galaxy sample. The signal-to-noise of the full data vector without angular scale cuts is $\sim 30$; the part of the data vector used for cosmological inference has a signal-to-noise of $\sim 20$. The main reduction of the signal-to-noise comes from uncertainty in the modeling of nonlinear galaxy bias, which necessitates removal of the small-angle \nk{} correlation measurements. Compared to the DES Y1 analysis, the signal-to-noise increased by a factor of $\sim 2$ and we are no longer limited by contamination of tSZ in the CMB lensing map.

The joint analysis of these two cross-correlations results in the constraints $\Omega_{\rm m} = 0.272^{+0.032}_{-0.052}$; $S_{8} = 0.736^{+0.032}_{-0.028}$ ($\Omega_{\rm m} = 0.245^{+0.026}_{-0.044}$; $S_{8} = 0.734^{+0.035}_{-0.028}$) when assuming linear (nonlinear) galaxy bias in our modeling. For $S_{8},$ these constraints are more than a factor of 2 tighter than our DES Y1 results, $\sim$30\% looser than constraints from DES Y3 cosmic shear and $\sim$70\% looser than constraints from DES Y3 \threetwo{}. We highlight here several interesting findings from this work:
\begin{itemize}
    \item We find that \gk{} dominates the constraints in the \nkgk{} combination, confirming our findings from the simulated analysis in \citetalias{y3-nkgkmethods}.
    \item We find that the lensing ratio has a large impact on the \nkgk{} constraints, improving the $S_8$ constraints by $\sim40\%$. In addition, the \nkgk{} data vector constrains the $\eta_1-\eta_2$ degeneracy direction, something not seen in the DES Y3 \threetwo{} data vectors.
    \item We investigate the use of two alternative lens samples for the analysis: the 6-bin \maglim{} sample and the \redmagic{} sample. In contrast to the fiducial DES Y3 \threetwo{} analysis, we find that the \nkgk{} analysis using the two alternative lens samples pass our unblinding criteria and show no signs of systematic contamination. 
    \item With fixed cosmology, we use the \nkgk{} data vector to constrain the galaxy bias values using the 6-bin \maglim{} sample and the \redmagic{} sample. For the two high-redshift \maglim{} bins, we find bias values that agree more with galaxy clustering. On the other hand, for the \redmagic{} sample, we find bias values more consistent with galaxy-galaxy lensing. These provide additional information for understanding the systematic effect seen in  \cite{y3-3x2ptkp} from these two alternative lens samples.
    \item Comparing with previous cosmic shear and \gk{} constraints, we find that in line with previous findings, our \gk{} constraint on $S_8$ is lower than the primary CMB constraint from {\it Planck}. In addition, for the first time, \gk{} has achieved comparable precision to state-of-the-art cosmic shear constraints.
\end{itemize}

The constraints derived in this paper from \nkgk{} can now be compared and combined with the DES Y3 \threetwo{} probes \citep{y3-3x2ptkp}, which we will do in \citetalias{y3-5x2kp}.  We will present therein our final combined results along with tests for consistency with external datasets.  It is however intriguing that with the galaxy-CMB lensing cross-correlation probes alone, our datasets provide very competitive constraints on the late-time large-scale structure compared to galaxy-only probes. Due to the relative insensitivity to certain systematic effects, this additional constraint is especially important for cross-checking and significantly improving the robustness of the galaxy-only results. Another unique aspect of this work compared to other cross-correlation analyses is that we have carried out our work in an analysis framework that is fully coherent with the galaxy-only probes, making it easy to compare and combine.

Looking forward to the final datasets from DES, SPT and ACT, as well as datasets from the Vera C. Rubin Observatory's Legacy Survey of Space and Time\footnote{\url{https://www.lsst.org}} (LSST), the ESA's Euclid mission\footnote{\url{https://www.euclid-ec.org}}, the Roman Space Telescope\footnote{\url{https://roman.gsfc.nasa.gov}}, the Simons Observatory\footnote{\url{https://simonsobservatory.org/}} (SO), and CMB Stage-4\footnote{\url{https://cmb-s4.org/}} (CMB-S4), our results show that there are significant opportunities for combining the galaxy and CMB lensing datasets to both improve the constraints on cosmological parameters and to make the constraints themselves more robust to systematic effects.

\acknowledgements

CC and YO are supported by DOE grant DE-SC0021949. 

The South Pole Telescope program is supported by
the National Science Foundation (NSF) through 
the grant OPP-1852617.  Partial support is also 
provided by the Kavli Institute of Cosmological Physics 
at the University of Chicago.
Argonne National Laboratory’s work was supported by
the U.S. Department of Energy, Office of Science, Office of High Energy Physics, under contract DE-AC02-
06CH11357. Work at Fermi National Accelerator Laboratory, a DOE-OS, HEP User Facility managed by the Fermi Research Alliance, LLC, was supported under Contract No. DE-AC02- 07CH11359. The Melbourne authors acknowledge support from the Australian Research Council’s Discovery Projects scheme (DP200101068). The McGill authors acknowledge funding from the Natural Sciences and Engineering Research
Council of Canada, Canadian Institute for Advanced research, and the Fonds de recherche du Qu\'{u}bec Nature et technologies. The CU Boulder group acknowledges support from NSF AST-0956135. The Munich group acknowledges the support by the ORIGINS Cluster (funded by the Deutsche Forschungsgemeinschaft (DFG, German Research Foundation) under Germany’s Excellence Strategy – EXC-2094 – 390783311), the MaxPlanck-Gesellschaft Faculty Fellowship Program, and
the Ludwig-Maximilians-Universit\"{a}t M\"{u}nchen. JV acknowledges support from the Sloan Foundation.

Funding for the DES Projects has been provided by the U.S. Department of Energy, the U.S. National Science Foundation, the Ministry of Science and Education of Spain, 
the Science and Technology Facilities Council of the United Kingdom, the Higher Education Funding Council for England, the National Center for Supercomputing 
Applications at the University of Illinois at Urbana-Champaign, the Kavli Institute of Cosmological Physics at the University of Chicago, 
the Center for Cosmology and Astro-Particle Physics at the Ohio State University,
the Mitchell Institute for Fundamental Physics and Astronomy at Texas A\&M University, Financiadora de Estudos e Projetos, 
Funda{\c c}{\~a}o Carlos Chagas Filho de Amparo {\`a} Pesquisa do Estado do Rio de Janeiro, Conselho Nacional de Desenvolvimento Cient{\'i}fico e Tecnol{\'o}gico and 
the Minist{\'e}rio da Ci{\^e}ncia, Tecnologia e Inova{\c c}{\~a}o, the Deutsche Forschungsgemeinschaft and the Collaborating Institutions in the Dark Energy Survey. 

The Collaborating Institutions are Argonne National Laboratory, the University of California at Santa Cruz, the University of Cambridge, Centro de Investigaciones Energ{\'e}ticas, 
Medioambientales y Tecnol{\'o}gicas-Madrid, the University of Chicago, University College London, the DES-Brazil Consortium, the University of Edinburgh, 
the Eidgen{\"o}ssische Technische Hochschule (ETH) Z{\"u}rich, 
Fermi National Accelerator Laboratory, the University of Illinois at Urbana-Champaign, the Institut de Ci{\`e}ncies de l'Espai (IEEC/CSIC), 
the Institut de F{\'i}sica d'Altes Energies, Lawrence Berkeley National Laboratory, the Ludwig-Maximilians Universit{\"a}t M{\"u}nchen and the associated Excellence Cluster Universe, 
the University of Michigan, NFS's NOIRLab, the University of Nottingham, The Ohio State University, the University of Pennsylvania, the University of Portsmouth, 
SLAC National Accelerator Laboratory, Stanford University, the University of Sussex, Texas A\&M University, and the OzDES Membership Consortium.

Based in part on observations at Cerro Tololo Inter-American Observatory at NSF's NOIRLab (NOIRLab Prop. ID 2012B-0001; PI: J. Frieman), which is managed by the Association of Universities for Research in Astronomy (AURA) under a cooperative agreement with the National Science Foundation.

The DES data management system is supported by the National Science Foundation under Grant Numbers AST-1138766 and AST-1536171.
The DES participants from Spanish institutions are partially supported by MICINN under grants ESP2017-89838, PGC2018-094773, PGC2018-102021, SEV-2016-0588, SEV-2016-0597, and MDM-2015-0509, some of which include ERDF funds from the European Union. IFAE is partially funded by the CERCA program of the Generalitat de Catalunya.
Research leading to these results has received funding from the European Research
Council under the European Union's Seventh Framework Program (FP7/2007-2013) including ERC grant agreements 240672, 291329, and 306478.
We  acknowledge support from the Brazilian Instituto Nacional de Ci\^encia
e Tecnologia (INCT) do e-Universo (CNPq grant 465376/2014-2).

This manuscript has been authored by Fermi Research Alliance, LLC under Contract No. DE-AC02-07CH11359 with the U.S. Department of Energy, Office of Science, Office of High Energy Physics.

\appendix

\begin{figure*} 
\begin{center}
\includegraphics[width=0.9\linewidth]{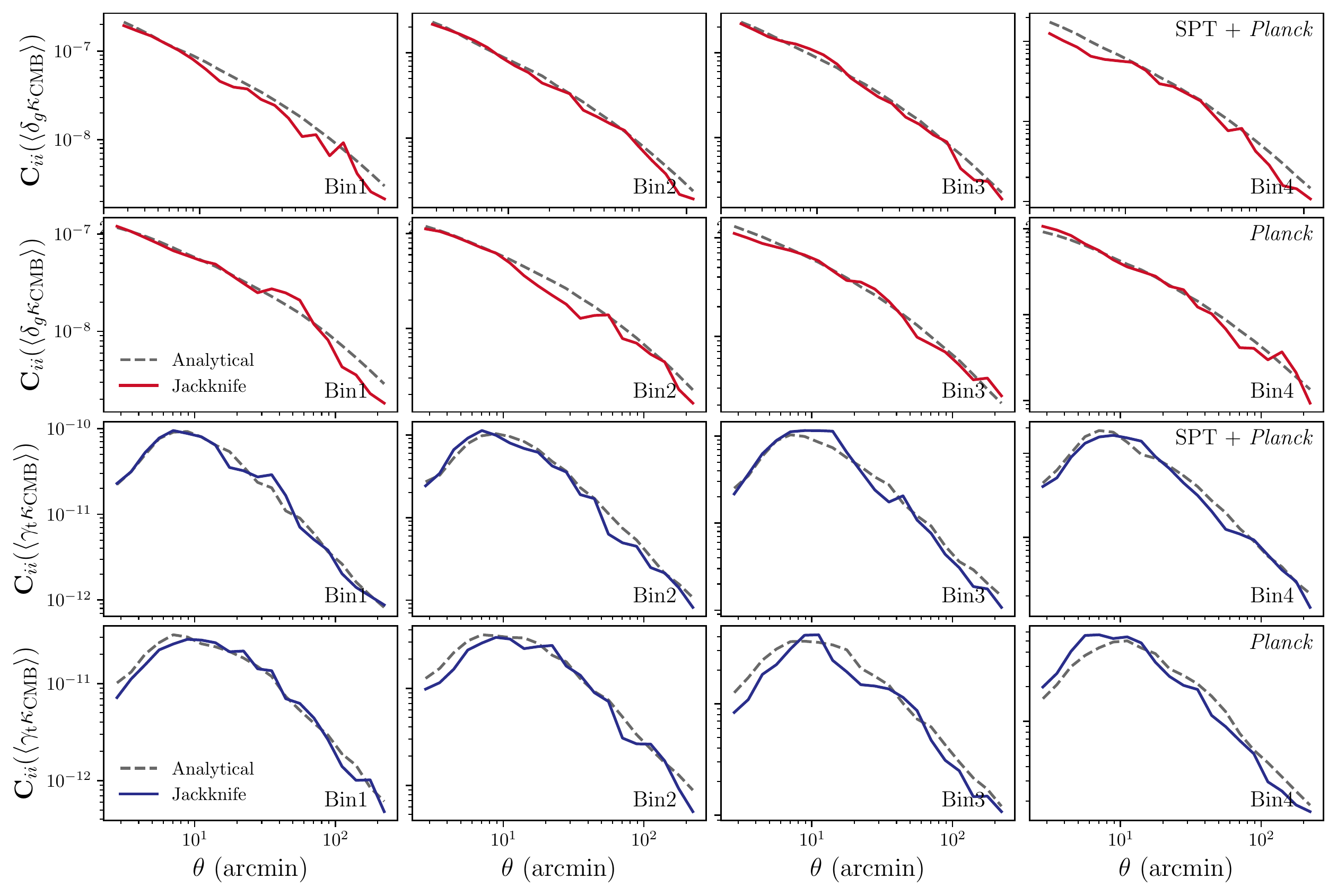}
\caption{Comparison between the diagonal elements of the jackknife covariance and our fiducial covariance matrix (analytical covariance with noise-noise correction applied).}
\label{fig:cov_jk_vs_analytical}
\end{center}
\end{figure*}

\section{Jackknife covariance matrix}
\label{sec:jk_cov}

We have performed extensive validation tests on our methodology of modeling in the covariance matrix in \citetalias{y3-nkgkmethods}. The ultimate check, however, is to compare the covariance matrix with a data-driven jackknife covariance matrix. The jackknife covariance incorporates naturally the noise in the data as well as any non-cosmological spatial variation in the data that might be important. This comparison was done after unblinding and the update of the covariance described in footnote 6, and is only used as a confirmation -- that is, we cannot change any analysis choices based on this check. 

In Figure~\ref{fig:cov_jk_vs_analytical} we show the diagonal elements of the jackknife covariance matrix (calculated using the delete-one block jackknife method by dividing the footprint into 80 patches) for the fiducial lens sample, compared with our fiducial covariance matrix. We find excellent agreement between them on all scales, both \nk{} and \gk{}, and on both the SPT+{\it Planck} and {\it Planck} patch.

\section{Diagnostic tests}
\label{sec:sys_test}

We perform a number of diagnostic tests to make sure that our measurements are not significantly contaminated by potential systematic effects. As we have discussed in Section~\ref{sec:intro}, cross-survey correlations like those presented here are expected to be inherently more robust to possible systematic effects. In addition, extensive tests have been done on both the galaxy and CMB data products in \cite{Omori:2017,y3-shapecatalog, y3-cosmicshear1, y3-cosmicshear2, y3-gglensing,y3-galaxyclustering}. We perform a series of diagnostic tests specific to the cross-correlation probes.

\begin{figure*}
\begin{center}
\includegraphics[width=0.95\linewidth]{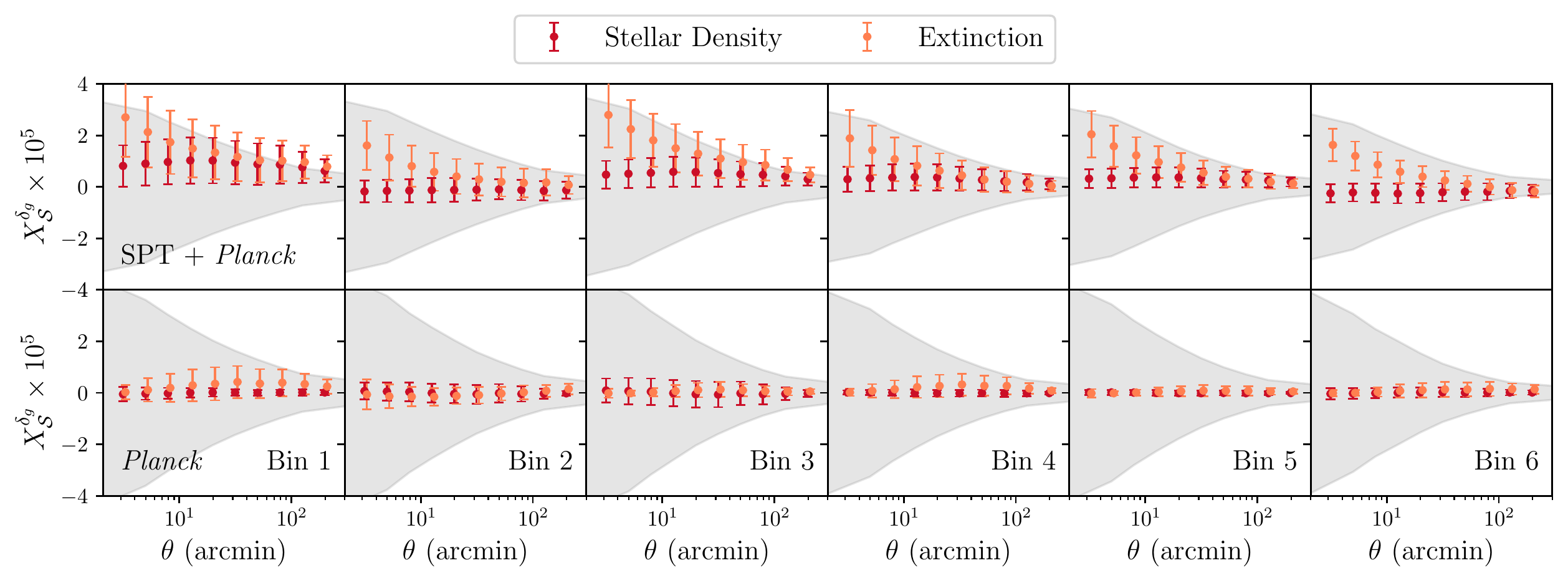}
\caption{
The measured systematic contamination of \nk{} for the \maglim{} lens sample, as assessed by Equation~\ref{eq:systematic_bias}, for the SPT+{\it Planck} field (top) and the {\it Planck} field (bottom) and for different redshift bins.  For reference, the grey band shows 10\% of the statistical uncertainties for the corresponding data vectors.  In all cases, the measured bias is significantly below the statistical uncertainties on the \nk{} measurements.}
\label{fig:sys_nk}
\end{center}
\end{figure*}

\begin{figure*} 
\begin{center}
\includegraphics[width=0.95\linewidth]{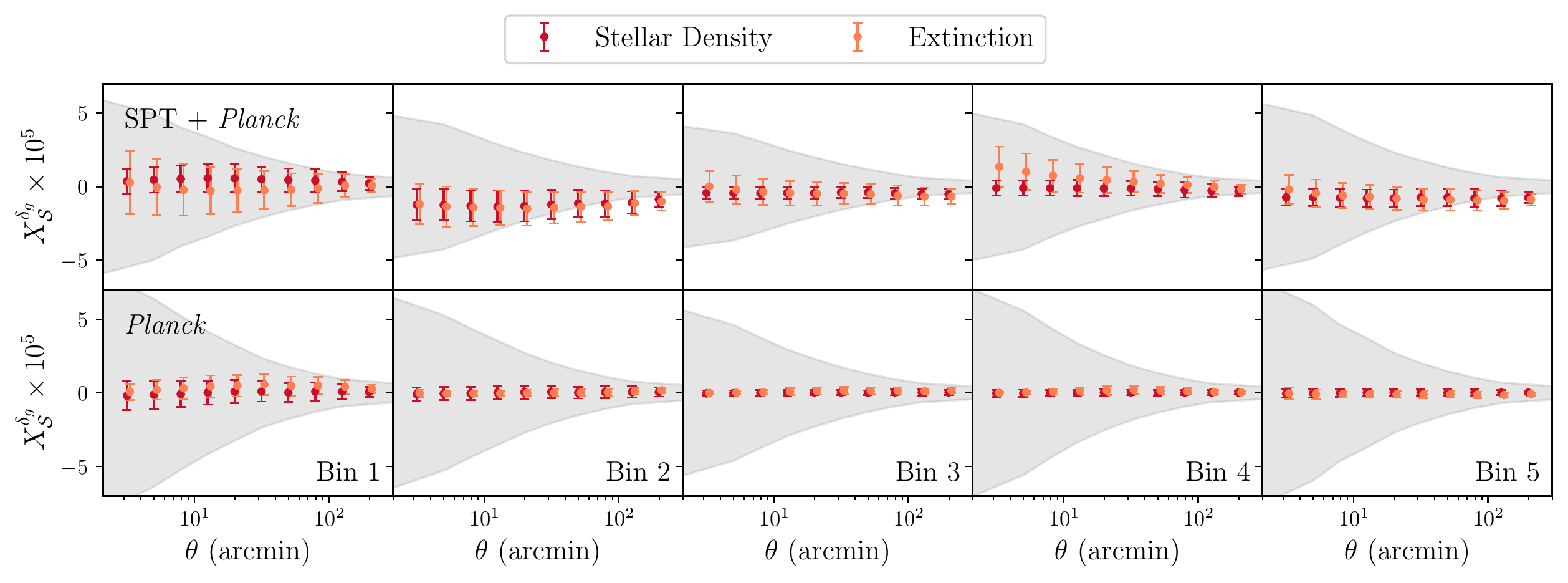}
\caption{Same as Figure~\ref{fig:sys_nk} but for the \redmagic{} lens sample.}
\label{fig:sys_nk2}
\end{center}
\end{figure*}

\begin{figure*} 
\begin{center}
\includegraphics[width=0.85\linewidth]{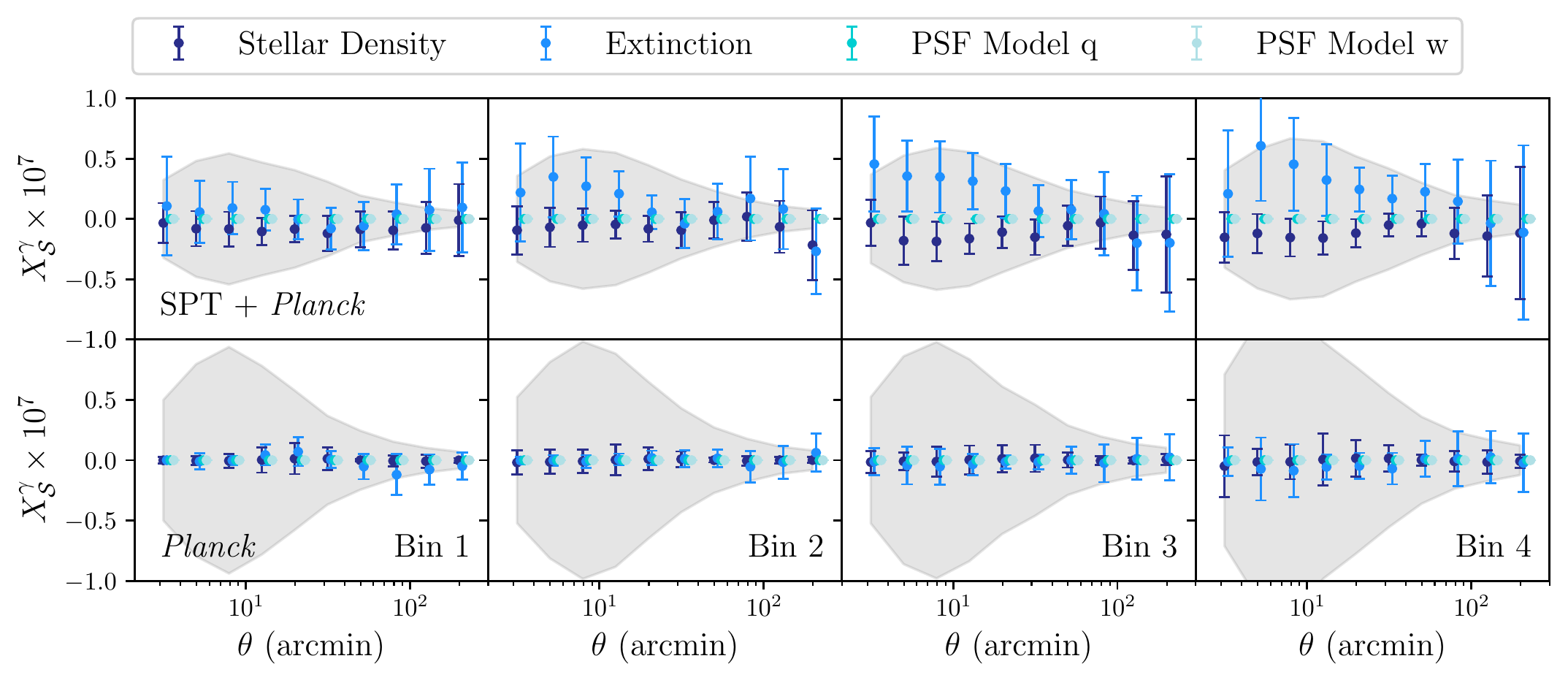}
\caption{The measured systematic contamination of \gk{},
as assessed by Equation~\ref{eq:systematic_bias}, for the SPT+{\it Planck} field (top) and the {\it Planck} field (bottom) and for different redshift bins. The grey band shows 1\% of the statistical uncertainties for the corresponding data vectors.} 
\label{fig:sys_gk}
\end{center}
\end{figure*}

\subsection{Cross-correlation with survey property maps}

If a given contaminant associated with some survey property simultaneously affects the galaxy and the CMB fields that we are cross-correlating, the cross-correlation signal will contain a spurious component that is not cosmological. A possible example is dust, which could simultaneously contaminate the CMB lensing map (through thermal emission in CMB bands) and the galaxy density field (through extinction).  In addition to dust, we consider several other possible survey properties. This test is designed to detect such effects. We calculate the correlation statistic, $X^f_{\mathcal{S}}$, between the observables of interest and various survey property maps:
\begin{equation}
\label{eq:systematic_bias}
X^{f}_{\mathcal{S}}(\theta)= \frac{\langle \kappa_{\rm CMB} \mathcal{S}  (\theta)\rangle \,  \langle f \mathcal{S}(\theta)\rangle} {\langle \mathcal{S} \mathcal{S}(\theta)\rangle},
\end{equation}
where $\mathcal{S}$ is the survey property map of interest, and $f$ is either $\delta_{g}$ or $\gamma_{\rm t}$. This expression captures correlation of the systematic with both $\kappa_{\rm CMB}$ and $f$, and is normalized to have the same units as $\langle f\kcmb \rangle$. Henceforth, we omit the $\theta$-dependence in the notation for simplicity, but note that all the factors in Equation~\ref{eq:systematic_bias} are functions of $\theta$. Unless the systematic map is correlated with both $f$ and $\kappa_{\rm CMB}$, it will not bias $\langle f\kcmb \rangle$ and $X^{f}_{\mathcal{S}}$ will be consistent with zero. Note that $X^{f}_{\mathcal{S}}$ should also be compared with the statistical uncertainty of $\langle f\kcmb \rangle$, as a certain systematic could be significantly detected but have little impact on the final results if it is much smaller than the statistical uncertainty.

For \nk{}, we consider two $\mathcal{S}$ fields: stellar density and extinction. For \gk{}, we look in addition at two fields associated with PSF modeling errors. The quantities $q$ and $w$ measure the point-spread function (PSF) modeling residuals as introduced in \cite{y3-shapecatalog}, $q=e_{*}-e_{\rm model}$ is the difference of the true ellipticity of the PSF as measured by stars and that inferred by the PSF model, and $w=e_{*}(T_{*}-T_{\rm model})/T_{*}$, where $T$ is a measure of size of the PSF, is the impact on the PSF model ellipticity when the PSF size is wrong by $T_{*}-T_{\rm model}$. As both $q$ and $w$ are spin-2 quantities like the ellipticity, we first decompose them into $E$ and $B$ modes using the same method used for generating weak lensing convergence maps in \cite{y3-massmapping}. We then use the $E$-mode maps as the $\mathcal{S}$ maps to perform the cross-correlation test. The rationale here is that if there is a non-trivial $E$-mode component, it could signify contamination in the shear signal and will correlate with the shear field.  

Figures~\ref{fig:sys_nk}, ~\ref{fig:sys_nk2} and \ref{fig:sys_gk} show the result of our measured $X^{f}_{\mathcal{S}}$ for the different parts of the data vector. For comparison, we also plot the statistical uncertainty on the data vector; given that the statistical uncertainties are much larger than the measured biases in all cases, we scale the statistical uncertainties by 0.1 (\nk{}) and 0.01 (\gk{}). The $\chi^2$ values per degree of freedom for the $X^{f}_{\mathcal{S}}$ measurements with respect to the null model are shown in Tables~\ref{table:sys_NK_ML},~\ref{table:sys_NK_RM} and~\ref{table:sys_GK} together with the probability-to-exceed (PTE) values. The $\chi^2$ as well as the error bars on the plots are derived from jackknife resampling where we use 65 equal-area jackknife patches for the SPT+{\it Planck} footprint and 85 patches for the {\it Planck} area. To obtain a more reliable jackknife covariance, we measure $X^{f}_{\mathcal{S}}$ using 10 angular bins instead of the 20 bins used for the data vectors. In general, most of the systematic effects are very consistent with zero.

\begin{table}
\centering
 \begin{tabular}{llcc}
\hline
&$\mathcal{S}$  & Stellar density & Extinction  \\  
&Bin & \multicolumn{2}{c}{$\chi^{2}$/dof (PTE)} \\
\hline
\parbox[t]{2mm}{\multirow{6}{*}{\rotatebox[origin=c]{90}{SPT+{\it Planck}}}} & 1 & 0.42 (0.85) & 0.90 (0.49) \\ 
&2 & 0.10 (0.99) & 0.65 (0.71) \\ 
&3 & 0.21 (0.98) & 0.64 (0.72) \\ 
&4 & 0.13 (0.99) & 1.12 (0.34) \\ 
&5 & 0.22 (0.98) & 1.34 (0.21) \\ 
&6 & 0.36 (0.93) & 1.66 (0.10) \\ 
\hline
\parbox[t]{2mm}{\multirow{6}{*}{\rotatebox[origin=c]{90}{{\it Planck}}}} & 1 & 0.02 (0.99) & 0.40 (0.87) \\ 
&2 & 0.12 (0.99) & 0.26 (0.96) \\ 
&3 & 0.15 (0.99) & 0.28 (0.96) \\ 
&4 & 0.08 (0.99) & 0.33 (0.93) \\ 
&5 & 0.06 (0.99) & 0.21 (0.98) \\ 
&6 & 0.05 (0.99) & 0.18 (0.98) \\ 
\hline
\end{tabular}
\caption{The $\chi^2$ per degree of freedom for the systematics diagnostics quantity (Equation~\ref{eq:systematic_bias}) for the \maglim{} \nk{} measurements. The different columns represent the different survey properties $\mathcal{S}$, whereas the different rows are for the tomographic bins in both the SPT+{\it Planck} patch and the {\it Planck} patch. The corresponding PTE values are listed in the parentheses.} 
\label{table:sys_NK_ML}
\end{table}

\begin{table}
\centering
 \begin{tabular}{llcc}
\hline
& $\mathcal{S}$  & Stellar density & Extinction  \\  
& Bin & \multicolumn{2}{c}{$\chi^{2}$/dof (PTE)} \\
\hline
\parbox[t]{2mm}{\multirow{5}{*}{\rotatebox[origin=c]{90}{SPT+{\it Planck}}}} & 1 & 0.09 (0.99) & 0.20 (0.97) \\ 
&2 & 0.50 (0.83) & 0.56 (0.78) \\ 
&3 & 0.42 (0.88) & 0.38 (0.91) \\ 
&4 & 0.28 (0.96) & 0.76 (0.62) \\ 
&5 & 0.73 (0.64) & 1.13 (0.33) \\  
\hline
\parbox[t]{2mm}{\multirow{5}{*}{\rotatebox[origin=c]{90}{{\it Planck}}}} & 1 & 0.09 (0.99) & 0.38 (0.89) \\ 
&2 & 0.09 (0.99) & 0.16 (0.99) \\ 
&3 & 0.05 (0.99) & 0.19 (0.98) \\ 
&4 & 0.04 (0.99) & 0.16 (0.99) \\ 
&5 & 0.04 (0.99) & 0.16 (0.99) \\ 
\hline
\end{tabular}
\caption{Same as Table~\ref{table:sys_NK_ML} but for the \redmagic{} lens sample.} 
\label{table:sys_NK_RM}
\end{table}

\begin{table*}
\centering
 \begin{tabular}{llccccc}
\hline
&$\mathcal{S}$    & Stellar density & Extinction & PSF model error $q$ & PSF model error $w$ & $\gamma_{\times}$  \\  
&Bin & \multicolumn{5}{c}{$\chi^{2}$/dof (PTE)} \\ \hline
\parbox[t]{2mm}{\multirow{5}{*}{\rotatebox[origin=c]{90}{SPT+{\it Planck}}}}&&&&&&\\
 & 1 & 0.12 (0.99) & 0.12 (0.99) & 0.34 (0.96) & 0.15 (0.99) & 1.11 (0.34)  \\ 
&2 & 0.17 (0.99) & 0.38 (0.95) & 0.20 (0.99) & 0.18 (0.99) & 1.18 (0.29)  \\ 
&3 & 0.32 (0.94) & 0.39 (0.90) & 0.40 (0.89) & 0.30 (0.95) & 0.60 (0.75)  \\ 
&4 & 0.20 (0.97) & 0.41 (0.86) & 0.19 (0.97) & 0.15 (0.98) & 1.91 (0.07)  \\ 
\hline
\parbox[t]{2mm}{\multirow{4}{*}{\rotatebox[origin=c]{90}{{\it Planck}}}} & 1 & 0.09 (0.99) & 0.06 (0.99) & 0.11 (0.99) & 0.08 (0.99) & 1.15 (0.31) \\ 
& 2 & 0.09 (0.99) & 0.04 (0.99) & 0.25 (0.98) & 0.17 (0.99) & 1.28 (0.23) \\ 
& 3 & 0.12 (0.99) & 0.07 (0.99) & 0.19 (0.99) & 0.14 (0.99) & 1.16 (0.31) \\ 
& 4 & 0.16 (0.99) & 0.18 (0.99) & 0.27 (0.98) & 0.18 (0.99) & 1.12 (0.33) \\ 
\hline

\end{tabular}
\caption{The $\chi^2$ per degree of freedom for the systematics diagnostics quantity (Equation~\ref{eq:systematic_bias}) for the \gk{} measurements. The different columns represent the different survey properties $\mathcal{S}$, whereas the different rows are for the tomographic bins in both the SPT+{\it Planck} patch and the {\it Planck} patch. The corresponding PTE values are listed in the parentheses. The last column lists the corresponding numbers for the cross-shear measurement described in Section~\ref{sec:cross-shear}.} 
\label{table:sys_GK}
\end{table*}

For \nk{}, we find that the absolute level of the potential systematic effects as quantified by $X^{f}_{\mathcal{S}}$ is 1-2 orders of magnitudes smaller than the statistical errors. There appears to be more cross-correlation for the SPT+{\it Planck} area, especially with extinction. All of the PTE values of these cross-correlations are above our threshold for concern of 0.01, so we deem these results acceptable. For \gk{}, we find that the absolute levels of the $X^{f}_{\mathcal{S}}$ measurements is much lower ($>2$ orders of magnitude) -- this is expected as it is much less obvious how the survey property maps will leave an imprint on the shear field. Interestingly, we also find that overall the error bars are larger in the SPT+{\it Planck} patch compared to the {\it Planck} patch. This can be due to the survey property maps containing higher spatial fluctuation in the SPT+{\it Planck} area as part of the footprint is close to the galactic plane or the Large Magellanic Cloud (LMC).

\begin{figure*} 
\begin{center}
\includegraphics[width=0.85\linewidth]{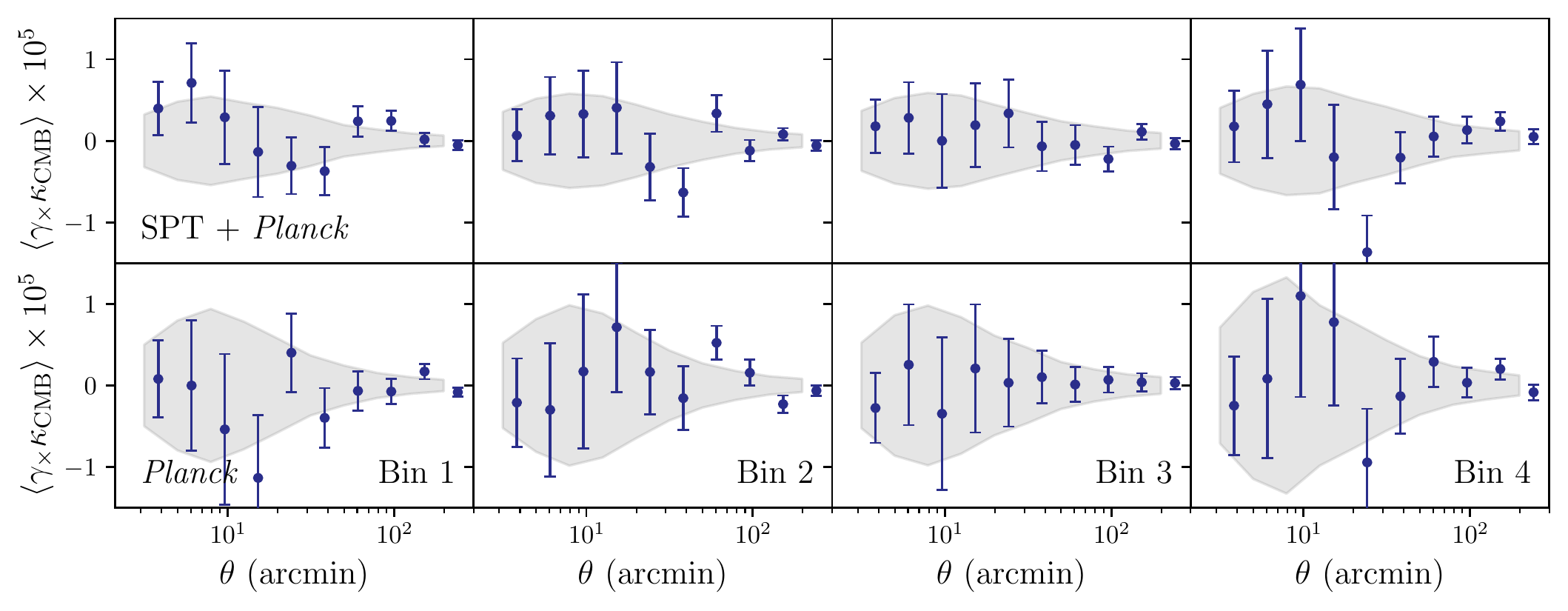}
\caption{Cross-correlation between than cross-component of shear with CMB lensing for the SPT+{\it Planck} field (top) and the {\it Planck} field (bottom) and for different redshift bins. The grey band shows the statistical uncertainties for \gk{}.} 
\label{fig:sys_gk2}
\end{center}
\end{figure*}

\subsection{Cross-shear component}
\label{sec:cross-shear}

During the measurement of \gk{}, we additionally measure its cross-shear counterpart $\langle \gamma_{\times}\kappa_{\rm CMB} \rangle$.  We replace $e_{t}$ in Equation~\ref{eq:measurement} with $e_{\times}$, the corrected ellipticity oriented 45$^{\circ}$ to the line connecting map pixel and the source galaxy. The correlation 
$\langle \gamma_{\times}\kappa_{\rm CMB} \rangle$ should be consistent with zero. Any significant detection of $\langle \gamma_{\times}\kappa_{\rm CMB} \rangle$ could signal systematic effects in the \gk{} measurements. 

Our results are shown in Figure~\ref{fig:sys_gk2} with the $\chi^2$ per degree of freedom and PTE values listed in Table~\ref{table:sys_GK}. We find no significant detection of $\langle \gamma_{\times}\kappa_{\rm CMB} \rangle$ in all parts of the data vector.

\subsection{\nk{} measurements with and without weights}

As discussed in \cite{y3-galaxyclustering}, weights are applied to the lens galaxies in order to remove correlations with various survey properties. When performing the \nk{} measurement in Equation~\ref{eq:nk1}, these weights are applied (i.e. the $\eta^{\delta_g}$). In a cross-correlation, the effect of these weights will be non-negligible if the systematic effect that is being corrected by the weights also correlates with the CMB lensing map. We note that this test is not always a null-test, as we consider it more correct to use the weights. Rather, it shows qualitatively the level of the correction from these weights -- naively, the smaller the correction to start with, the less likely the residual contamination will be.  

\begin{figure} 
\begin{center}
 \includegraphics[width=0.99\linewidth]{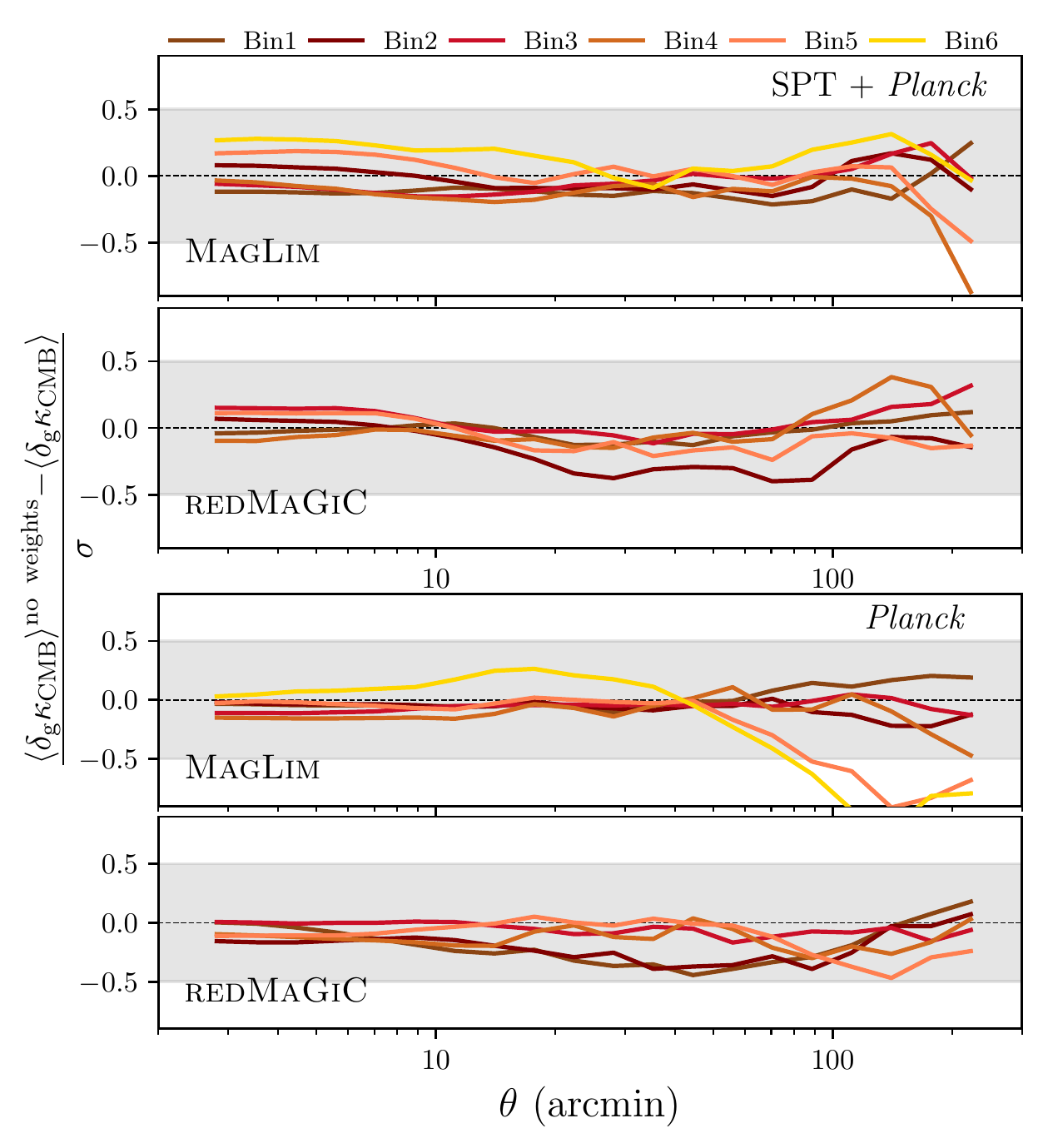}
\caption{The difference in the \nk{} cross-correlation between the two lens galaxy samples \textsc{MagLim} and \textsc{redMaGiC} with CMB lensing when using weights and without weights, over the statistical uncertainty of the measurement $\sigma$.   }
\label{fig:lss_weights}
\end{center}
\end{figure}

In Figure~\ref{fig:lss_weights} we show the difference between the \nk{} measurements with and without using the lens weights, for the two lens samples. To understand the significance of these results, we calculate the $\Delta \chi^2$ between the data vectors with and without weights for the fiducial \maglim{} sample, using the analytic covariance for the data vector and find a $\Delta \chi^2$ of 1.23 after scale cuts. Propagating this into cosmological constraints by running two chains using \nk{} with and without weights (fixing galaxy bias) gives a negligible $0.02 \sigma$ shift in the $\Omega_{\rm m}-S_{8}$ plane. It is also worth pointing out that we see that the weights most significantly affect the two high-redshift bins in the \maglim{} sample, this is likely due to the fact that the high-redshift bins are fainter and more affected by the spatially varying observing conditions.

\subsection{Biases from source masking}

In constructing the CMB lensing maps for this analysis, we apply a special procedure at the locations of bright point sources to reduce their impact on the output lensing maps.  As described in more detail in \citetalias{y3-nkgkmethods}, the CMB lensing estimator that we use involves two CMB maps, or ``legs.''  One of these is high-resolution map (i.e. the SPT+ {\it Planck} temperature map), and the other is a low-resolution tSZ-cleaned map (i.e. the {\it Planck} SMICAnosz temperature map).  To reduce the impact of point sources, we inpaint the point sources with fluxes  6.4$<$$F$$<$200~mJy using the method described in \cite{benoitlevy2013}.  The total inpainted area is roughly 3.6\% of the map.  The corresponding location in the tSZ-cleaned map are left untouched. We expect this procedure to result in a reasonable estimate of $\kappa_{\rm CMB}$ at the locations of the point sources, given that only one leg is inpainted, and the area being inpainted is small (such that Gaussian constrained inpainting predicts the pixels values of the inpainted region well) although it is possible that the noise properties of these regions differ somewhat from the map as a whole.  

\begin{figure} 
\begin{center}
 \includegraphics[width=0.99\linewidth]{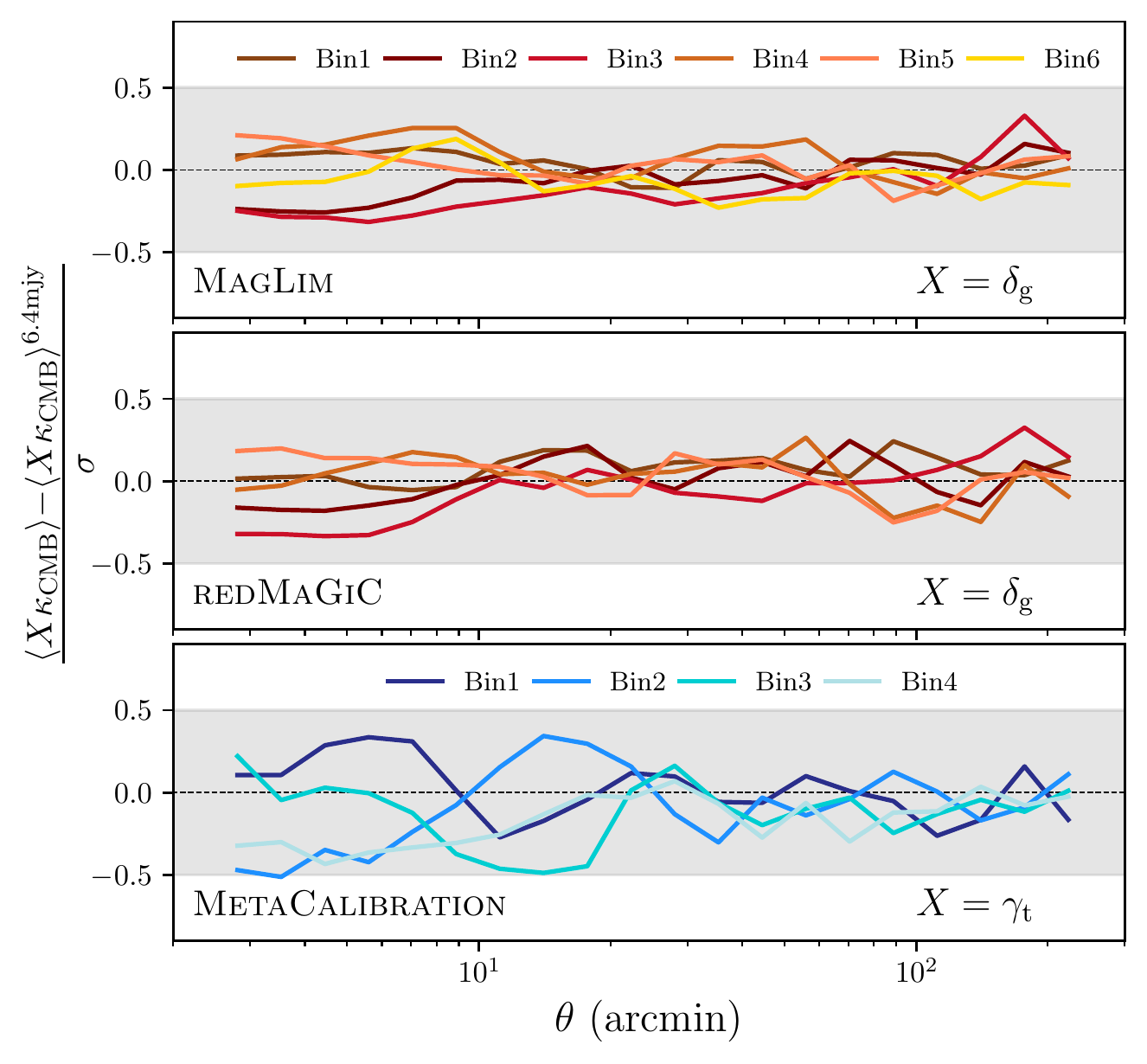}
\caption{Difference in the data vectors using the alternative mask and the fiducial one. This test is only done for the SPT + {\it Planck} patch, as it is specific to the SPT lensing reconstruction. 
}
\label{fig:test_mask}
\end{center}
\end{figure}

To test whether the inpainting procedure results in any bias, we also measure the cross-correlation with the lensing map after masking (i.e. completely removing) all the point sources down to 6.4~mJy. We show in Figure~\ref{fig:test_mask} the difference in the data vectors using the alternative mask and the fiducial one.   We find that there is no coherent difference in the correlation measurements across the range of angular scales considered.  There is, however, some scatter about our nominal measurements.  The level of this scatter is small, roughly $0.25$ and  $ 0.50\sigma$  across the full range of angular scales for \nk{} and \gk{} respectively.\footnote{This scatter results from the slightly higher-noise region caused by the half-leg lensing reconstruction, with the point sources left in the non-inpainted map effectively behaving as noise.}
Given that such scatter is expected to have negligible impact on our results, and since some scatter between the is expected simply due to the different selection of pixels in the masked and unmasked CMB lensing maps, we do not find this to be a cause for worry.  Our baseline results will use the unmasked version of the CMB lensing map.

\begin{figure} 
\begin{center}
 \includegraphics[width=0.99\linewidth]{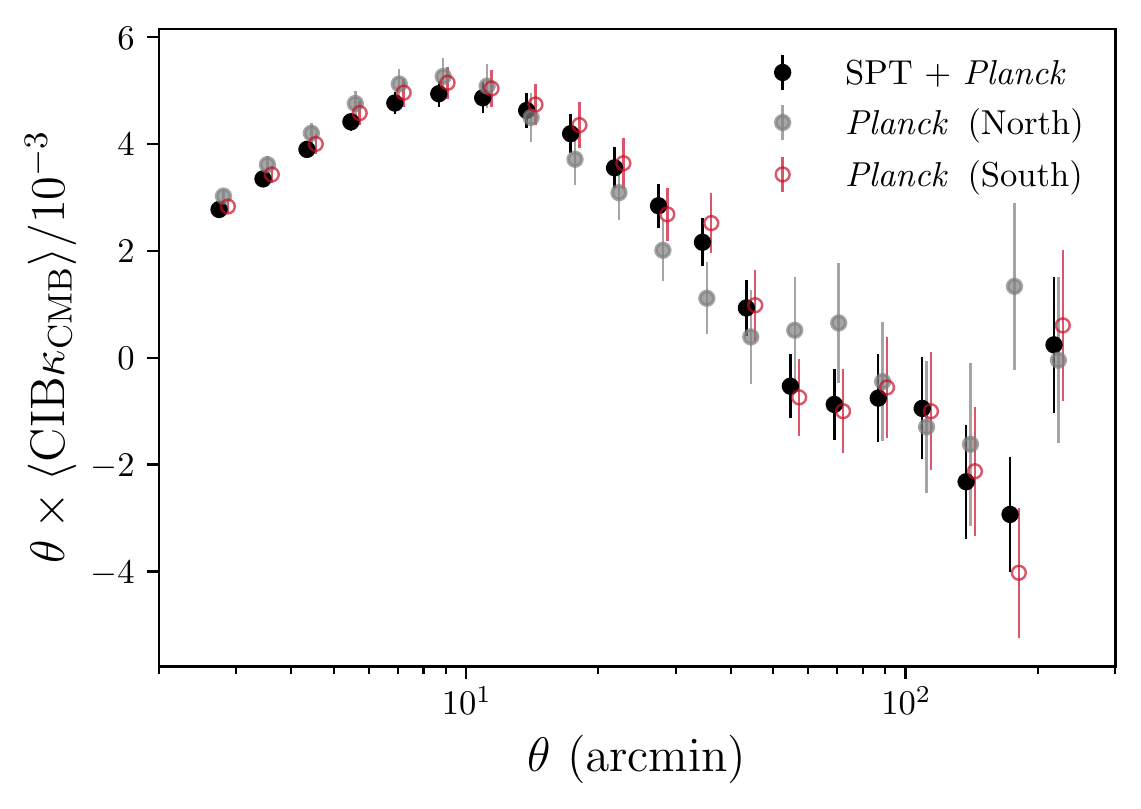}
\caption{Cross-correlation between CIB and the {\it Planck} lensing map in the North patch (solid grey), the {\it Planck} lensing map in the South patch (open red), and the SPT+{\it Planck} lensing map (black).}
\label{fig:CIB_xcorr}
\end{center}
\end{figure}

\begin{figure*}
\begin{center}
\includegraphics[width=1.0\linewidth]{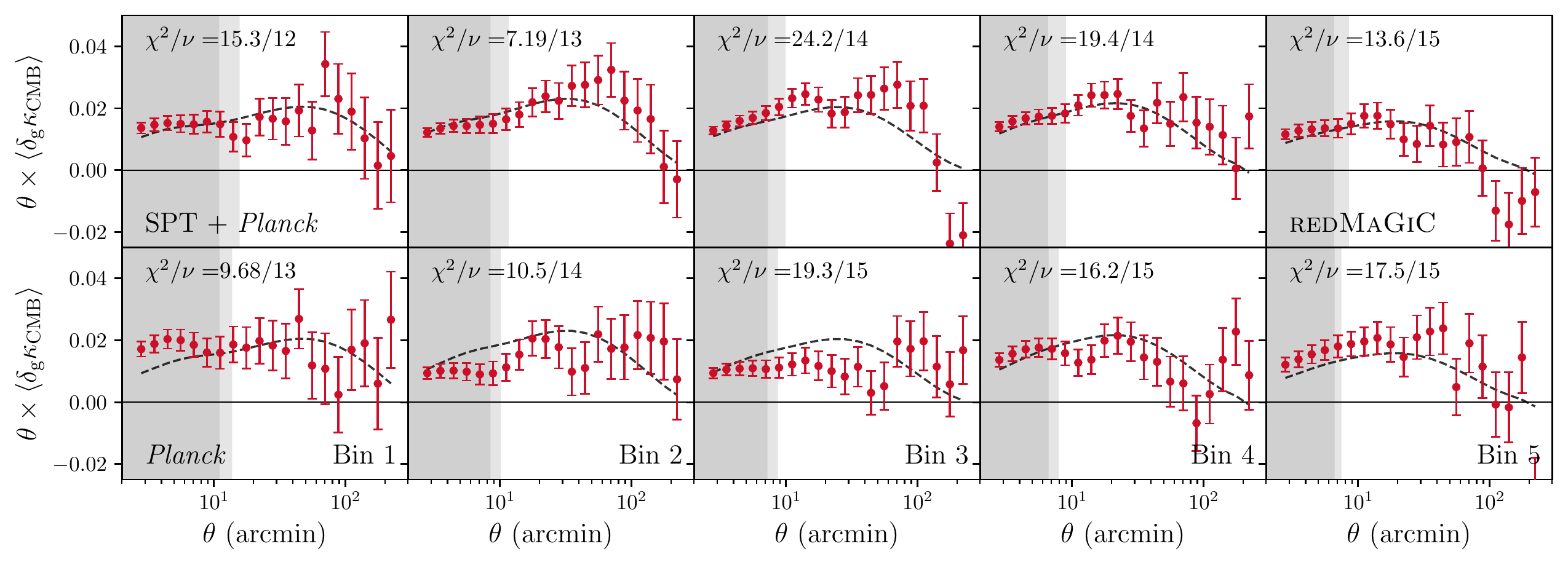}
\caption{Same as Figure~\ref{fig:davtavec_maglim} but for the \redmagic{} sample.}
\label{fig:wnk}
\end{center}
\end{figure*}

\subsection{Variations in the CMB lensing map}
Our fiducial analysis uses the SPT+{\it Planck} map in the Dec.$<-40^{\circ}$ region and the {\it Planck} lensing map in the region Dec.$>-39.5^{\circ}$. We left a 0.5$^{\circ}$ gap between the two maps to avoid correlation between the large-scale structure on the boundary. In order to verify that the cross-correlation with another LSS tracer is consistent between the two patches and two lensing data sets we (1) compare the cross-correlations between an external tracer of large-scale structure and the {\it Planck} lensing map split into two sub-regions (the ``North'' region with DEC$>-39.5^{\circ}$ and the ``South'' region with DEC$<-40^{\circ}$), and verify that that they are consistent, and (2) compare the cross-correlations between an external tracer of large-scale structure the {\it Planck} CMB lensing map in the South patch and the cross-correlation between the same external tracer with the SPT+{\it Planck} lensing map over the same sky area to test for consistency. As the external tracer of large-scale structure, we choose to use the CIB map from \cite{Lenz2019}.\footnote{Here we use the $n_{H}=2.5e20{\rm cm}^{-1}$ maps as defined in \protect\cite{Lenz2019}.} 

The resulting correlation measurements are shown in Figure~\ref{fig:CIB_xcorr} -- the high signal-to-noise is expected due to the significant overlap in the kernels of the two tracers. We make two comparisons:
\begin{enumerate}
\item CIB $\times$ {\it Planck} North vs. CIB $\times$ {\it Planck} South: We find a two-sample $\chi^{2}/\nu$ of $24.28/20$ with a PTE of 0.23. This demonstrates that the two patches are consistent with each other.

\item SPT+{\it Planck} vs. {\it Planck} South:
We compute the two-sample $\chi^{2}$, and find $\chi^{2}/\nu=23.9/20.$, with a PTE of 0.25. This demonstrates that the two measurements are consistent with each other.
\end{enumerate}

We note that there are two caveats associated with these cross-correlation measurements. The first is that, at 545 GHz, galactic emission is non-negligible, and while the CIB maps from \cite{Lenz2019} are intended to be free of galactic dust, there may be residuals. Second, the CIB-$\kappa_{\rm CMB}$ correlation is most sensitive to redshifts higher than those probed by DES galaxies. Still, it seems unlikely the $\kappa_{\rm CMB}$ maps could have spatially varying biases that correlate with low redshift structure if the CIB-$\kappa$ correlation does not show such biases.

\section{\redmagic{} results}
\label{sec:redmagic}

In this appendix we show the results for the second lens sample -- the \redmagic{} sample. The data vector is shown in Figure~\ref{fig:wnk} with signal-to-noise values listed in Table~\ref{table:s2n}. We find that (1) no significant systematic effects were found as described in Appendix~\ref{sec:sys_test}, (2) we get a $p$-value greater than 0.01 when comparing the \nkgk{} constraints from {\it Planck} to constraints from SPT+{\it Planck}, and (3) the goodness-of-fit of the fiducial \nkgk{} unblinded chain corresponds to a $p$-value greater than 0.01. These results allowed us to unblind our results, and the final constraints are listed in Table~\ref{table:cosmo} and the fiducial constraints are shown in Figure~\ref{fig:maglimvsredmagic}. 

%\bibliography{main.bbl}
\bibliography{bibl,y3kp}

\end{document}